\documentstyle[aps,prl]{revtex}
\begin{document}
\title{The shape of a moving fluxon in stacked Josephson junctions.}
\author{V.M.Krasnov$^{a,b}$, and D.Winkler$^a$.}
\address{a) Department of Microelectronics and Nanoscience, Chalmers University of\\
Technology, S-41296 G\"oteborg, Sweden \\
b) Institute of Solid State Physics, 142432 Chernogolovka, Russia }
\date{\today }
\maketitle

\begin{abstract}
We study numerically and analytically the shape of a single fluxon moving in
a double stacked Josephson junctions (SJJ's) for various junction
parameters. We show that the fluxon in a double SJJ's consists of two
components, which are characterized by different Swihart velocities and
Josephson penetration depths. The weight coefficients of the two components
depend on the parameters of the junctions and the velocity of the fluxon. It
is shown that the fluxon in SJJ's may have an unusual shape with an inverted
magnetic field in the second junction when the velocity of the fluxon is
approaching the lower Swihart velocity. Finally, we study the influence of
fluxon shape on flux-flow current-voltage characteristics and analyze the
spectrum of Cherenkov radiation for fluxon velocity above the lower Swihart
velocity. Analytic expression for the wavelength of Cherenkov radiation is
derived.
\end{abstract}


I. INTRODUCTION

Properties of stacked Josephson junctions (SJJ's) are of considerable
interest both for applications in cryoelectronics and for fundamental
physics. A particular interest in SJJ's was stimulated by the discovery of
high-$T_c$ superconductors (HTSC). Highly anisotropic HTSC compounds, such
as Bi$_2$Sr$_2$CaCu$_2$O$_{8+x}$, may be considered as stacks of atomic
scale intrinsic Josephson junctions\cite{Muler}. The layered structure
determines many of the unusual properties of HTSC. Behavior of model low-$%
T_c $ SJJ's and HTSC exhibit many similarities\cite{Klein2}. Due to mutual
coupling of the junctions in the stack, the physical properties of SJJ's can
be qualitatively different from those of single Josephson junctions (JJ's).
Therefore, a one-to-one comparison between single and stacked Josephson
junctions is difficult to do. Hence, the basic properties of SJJ's have to
be studied in order to describe correctly the Josephson behavior of layered
superconductors.

Perpendicular ($c$-axis) transport measurements in magnetic field, $H$,
parallel to layers ($ab$-plane) is an explicit way of studying Josephson
phenomena in SJJ's. In this case the magnetic field penetrates the stack in
the form of Josephson-type vortices (fluxons), and the {\it c}-axis voltage
is caused by motion of such vortices along the layers. The shape of a fluxon
in SJJ's is different both from that of an Abricosov vortex in bulk
superconductor, since it does not have a normal core, and from that of
single JJ, since the circulating currents are not confined within one
junction. The behavior of SJJ's becomes particularly complicated when the
length of the stack in one direction, $L$, is much longer than the Josephson
penetration depth, $\lambda _J$. One of the unusual properties of long SJJ's
is the existence of multiple quasi-equilibrium fluxon modes\cite{Modes}, and
submodes\cite{Submod}, which are characterized by different fluxon
configurations in the stack. Due to the existence of such modes/submodes,
the state of the stack is not well defined by external conditions. Rather it
can be described only statistically with a certain probability of being in
any of the quasi-equilibrium states. Experimental evidences for the
existence of such modes were obtained both for HTSC intrinsic SJJ's\cite
{Submod,ISEC,Mros} and low-$T_c$ multilayers\cite{NbCu1}. The existence of
fluxon modes/submodes dramatically changes the behavior of long strongly
coupled SJJ's with respect to that of single long JJ's. An example of this
is the critical current, $I_c$, which becomes multiple valued\cite
{Submod,ISEC,Mros}; the fluctuations of $I_c$ become anomalously large\cite
{Submod,Mros}; and the magnetic field dependence of $I_c$ becomes very
complicated without periodicity in $H$\cite{Submod}.

For understanding both the static and dynamic properties of SJJ's, the shape
of the fluxon in SJJ's is important, and should be determined. In the static
case, the shape of the single fluxon was studied for layered superconductors
consisting of an infinite number of thin identical\cite{Clem} or nonidentical%
\cite{Kr-Gol} layers and for SJJ's \cite{Modes,SBP}. In our previous work%
\cite{Modes}, we have shown that in double SJJ's, two special single
component fluxon solutions exist, which are characterized by different
Swihart velocities and Josephson penetration depths. An approximate analytic
fluxon solution was suggested as a linear combination of the single
component solutions. For the static case, the approximate solution was shown
to be in a quantitative agreement with numerically obtained solutions.
Extending the approximate analytic solution to the dynamic case, it was
shown that drastic changes in the fluxon shape could occur with increasing
the fluxon velocity, resulting e.g. in possible inversion of the sign of the
magnetic field in the second junction and appearance of attractive fluxon
interaction\cite{Modes}. On the other hand, the choice between the special
single component solutions and the approximate analytic fluxon solution was
not addressed and the dependence of the fluxon shape on the junction
parameters was not studied. Using the perturbation approach, the second
order correction to the approximate analytic solution was derived and the
accuracy of the solution was analyzed in the recent paper\cite{Fluxon}.

To our knowledge, no comprehensive analysis of the single fluxon shape in
SJJ's exists for the dynamic case. The scope of the current paper is to
study quantitatively the shape of the moving fluxon in double SJJ's for
various junction parameters. Our analysis is based on numerical simulations
and analytical treatments of the coupled sine-Gordon equation, which
describes the physical properties of SJJ's\cite{SBP}. We show that the
single moving fluxon in double SJJ's may be described by both a single
component solution and a double component solution depending on the
parameters of the stack and the fluxon velocity. Moreover, the shape of the
fluxon may be quite anomalous, with inverted magnetic field in the second
junction and with nonmonotonous change of phase.

The paper is organized as follows: in section II, we rewrite the coupled
sine-Gordon equation for the case of solitonic fluxon motion and review
analytic single fluxon solutions obtained in Refs.\cite{Modes,Fluxon}. In
section III, we present the results of numerical simulations for
frictionless fluxon motion for different parameters of SJJ's, compare it
with analytical predictions of Ref.\cite{Modes}, we also formulate and
verify conditions for observation of different fluxon solutions. In section
IV, we discuss implementations of the fluxon shape in experimental
situation. In subsection IV.A, we study the influence of finite damping and
simulate current-voltage characteristics. Finally, in subsection IV.B we
consider the case of nonsolitonic fluxon motion with the propagation
velocity larger than the lower Swihart velocity. We have shown that such
fluxon motion is accompanied by plasma wave exitations and derive the
expression for the wavelength of such ''Cherenkov'' radiation.

II. GENERAL RELATIONS

We consider a double stack with the overlap geometry, consisting of
junctions 1 and 2 with the following parameters: $J_{ci}$ -the critical
current density, $C_i$ -the capacitance, $t_i$- the thickness of the tunnel
barrier between the layers, $d_i$ and $\lambda _{Si}$ - the thickness and
London penetration depth of superconducting layers and {\it L}- the length
of the stack. Hereafter the subscript $i$ on a quantity represents its
number. The elements of the stack are numerated from the bottom to the top,
so that junction $i$ consists of superconducting layers $i$, $i+1$ and the
tunnel barrier $i$. The fluxon will be placed in junction 1, if not stated
otherwise.

The physical properties of SJJ's are described by the coupled sine-Gordon
equation\cite{SBP}, which for a double stack with the overlap geometry can
be written as:

\begin{equation}
\left| 
\begin{array}{c}
\varphi _1^{\prime \prime } \\ 
\varphi _2^{\prime \prime }
\end{array}
\right| =\left| 
\begin{array}{cc}
1 & -S_2/\Lambda _1 \\ 
-S_2/\Lambda _1 & \Lambda _2/\Lambda _1
\end{array}
\right| \left| 
\begin{array}{c}
sin\left( \varphi _1\right) +\stackrel{\cdot \cdot }{\varphi _1}+\alpha _1%
\stackrel{\cdot }{\varphi _1}-\frac{J_b}{J_{c1}} \\ 
\frac{J_{c2}}{J_{c1}}sin\left( \varphi _2\right) +\frac{C_2}{C_1}\stackrel{%
\cdot \cdot }{\varphi _2}+\alpha _2\stackrel{\cdot }{\varphi _2}-\frac{J_b}{%
J_{c1}}
\end{array}
\right| ,  \label{Eq1}
\end{equation}

where $\varphi _{1,2}$ are gauge invariant phase differences in JJ's 1 and
2, 'prime' and 'dot' on the quantity represent partial derivatives in space
and time, respectively. Space and time are normalized to Josephson
penetration depth, $\lambda _{J1}=\sqrt{\frac{\Phi _0c}{8\pi ^2J_{c1}\Lambda
_1}}$, and inverted Josephson plasma frequency, $\omega _{p1}^{-1}=\sqrt{%
\frac{\Phi _0C_1}{2\pi cJ_{c1}}}$, respectively, of the single JJ 1. Here $%
\Phi _0$ is the flux quantum, $c$ is the velocity of light in vacuum and

$\Lambda _i=t_i+\lambda _{Si}coth\left( \frac{d_i}{\lambda _{Si}}\right)
+\lambda _{Si+1}coth\left( \frac{d_{i+1}}{\lambda _{Si+1}}\right) ,$

$S_i=\lambda _{Si}cosech\left( \frac{d_i}{\lambda _{Si}}\right) .$

The last terms in the right hand side of Eq.(1) represent total currents in
the JJ's, which consist of superconducting, displacement and quasiparticle
contributions, and $J_b$ represents bias current density. Viscous damping
due to quasiparticle current is characterized by the damping coefficient, $%
\alpha _i=\beta _{ci}^{-1/2}$, where $\beta _{ci}$ is the McCumber parameter
of single JJ $i$.

The coupling strength in the double SJJ's is described by a coupling
parameter

$S=\frac{S_2}{\sqrt{\Lambda _1\Lambda _2}}.$

The magnetic induction in the stack is equal to \cite{Modes}

\[
B_1=\frac{H_0}{2\left( 1-S^2\right) }\left[ \varphi _1^{\prime }+\frac{S_2}{%
\Lambda _1}\varphi _2^{\prime }\right] , 
\]

\begin{equation}
B_2=\frac{H_0}{2\left( 1-S^2\right) }\left[ \frac{S_2}{\Lambda _2}\varphi
_1^{\prime }+\frac{\Lambda _1}{\Lambda _2}\varphi _2^{\prime }\right] ,
\label{Eq2}
\end{equation}
where $H_0=\frac{\Phi _0}{\pi \lambda _{J1}\Lambda _1}$.

For the soliton-like fluxon motion, the phase differences in the stack
remain unchanged in the coordinate frame moving along with the fluxon.
Introducing the self-coordinate of the fluxon, $\xi =x-ut$, and neglecting
damping coefficient, we simplify Eq.(1) and rewrite it as a system of
coupled ordinary differential equations (ODE):

\[
\varphi _{1\xi \xi }^{\prime \prime }\left[ \frac{ab-S^2}{1-S^2}\right]
=asin\left( \varphi _1\right) -\frac{J_{c2}S_2}{J_{c1}\Lambda _1}sin\left(
\varphi _2\right) , 
\]

\begin{equation}
\varphi _{2\xi \xi }^{\prime \prime }\left[ \frac{ab-S^2}{1-S^2}\right] =b%
\frac{J_{c2}\Lambda _2}{J_{c1}\Lambda _1}sin\left( \varphi _2\right) -\frac{%
S_2}{\Lambda _1}sin\left( \varphi _1\right) ,  \label{Eq3}
\end{equation}
where 
\[
a=1-\frac{u^2}{c_{01}^2}\frac{C_2\Lambda _2}{C_1\Lambda _1}\left(
1-S^2\right) , 
\]

\begin{equation}
b=1-\frac{u^2}{c_{01}^2}\left( 1-S^2\right) ,  \label{Eq4}
\end{equation}
and $c_{01}=\lambda _{j1}\omega _{p1}$ is the Swihart velocity of the single
junction 1. Comparing Eqs. (1) and (3) it is seen that solution of the
coupled sine-Gordon equation for soliton-like fluxon motion is now reduced
to solution of the static problem, but with parameters depending on the
fluxon velocity.

Eq.(3) has a first integral,

\begin{equation}
\frac 1{1-S^2}\left[ b\frac{\left( \varphi _{1\xi }^{\prime }\right) ^2}2+a%
\frac{\Lambda _1}{\Lambda _2}\frac{\left( \varphi _{2\xi }^{\prime }\right)
^2}2+\frac{S_2}{\Lambda _2}\varphi _{1\xi }^{\prime }\varphi _{2\xi
}^{\prime }\right] +cos\left( \varphi _1\right) +\frac{J_{c2}}{J_{c1}}%
cos\left( \varphi _2\right) ={\bf C},  \label{Eq5}
\end{equation}
which reduces to that from Ref.\cite{Modes} for the static case, $u=0$. Here 
${\bf C}$ is a constant of the first integral.

{\it A. Special single component solutions}

In Ref.\cite{Modes} it was shown that Eq.(1), linearized with respect to $%
\varphi _2$, allows two special single component solutions of the type\cite
{asin}

\begin{eqnarray}
\varphi _1(\xi ) &=&F_{1,2}=4arctan\left[ exp\left( \xi /\lambda
_{1,2}\right) \right] ,  \label{Eq6} \\
\varphi _2 &=&arcsin\left( 1/\kappa _{1,2}sin\left( \varphi _1\right)
\right) ,  \nonumber
\end{eqnarray}
where $\kappa _{1,2}$ are solutions of the quadratic equation:

\begin{equation}
\frac{S_2}{\Lambda _1}\kappa ^2+\kappa \left[ 1-\frac{J_{c2}\Lambda _2}{%
J_{c1}\Lambda _1}+\frac{u^2}{c_{01}^2}\frac{\Lambda _2}{\Lambda _1}\left( 
\frac{J_{c2}}{J_{c1}}-\frac{C_2}{C_1}\right) \left( 1-S^2\right) \right] -%
\frac{J_{c2}S_2}{J_{c1}\Lambda _1}=0.  \label{Eq7}
\end{equation}
Therefore, for a double SJJ's there exist two characteristic Josephson
penetration depths,

\begin{equation}
\lambda _{1,2}^2=\frac{\lambda _{j1}^2}{1+\kappa _{2,1}S_2/\Lambda _1}\left(
1-\frac{u^2}{c_{1,2}^2}\right) ,  \label{Eq8}
\end{equation}
and two characteristic velocities

\begin{equation}
c_{1,2}^2=\frac{c_{01}^2}{1+\kappa _{2,1}\left( C_2J_{c1}S_2\right) /\left(
C_1J_{c2}\Lambda _1\right) }.  \label{Eq9}
\end{equation}

{\it B. Double component solution}

Taking the single component solutions as eigen-functions of the linearized
coupled sine-Gordon equation, an approximate analytic single fluxon solution
in JJ 1 was obtained in Ref.\cite{Modes}:

\[
\varphi _1=\frac{\kappa _1F_1-\kappa _2F_2}{\kappa _1-\kappa _2}, 
\]

\begin{equation}
\varphi _2=\frac{F_1-F_2}{\kappa _1-\kappa _2}.  \label{Eq10}
\end{equation}

Here $F_{1,2}$ are the single component solutions, Eq.(6). Recently this
solution was rederived more rigorously in Ref. \cite{Fluxon}. It was shown,
that for the static case Eq.(10) gives perfect approximation for $\varphi _1$
in the whole space region and for arbitrary parameters of the stack\cite
{Fluxon,Modes}. Using the perturbation approach the second order correction
to Eq.(10) was obtained in Ref.\cite{Fluxon}. As it is seen from Eq.(10) the
single fluxon in double SJJ's consists of two components. From Eq.(8) it is
seen that both components contract with increasing velocity, but the
characteristic velocities for the contraction are different for each
component and are given by Eq.(9). For identical junctions the contraction
of each component is of the Lorentz type, however, the contraction of the
fluxon itself is different from Lorentz contraction. This is a consequence
of the absence of Lorentz invariance for the coupled sine-Gordon equation.
For nonidentical junctions, the parameters $\kappa _{1,2}$, depend on the
fluxon velocity and thus contraction of the components is somewhat different
from Lorentz contraction. In this case the maximum characteristic velocity
should be obtained from the equation $u=c_{1,2}$ and Eqs. (7,9). By analogy
with single JJ's we'll refer the maximum characteristic velocities to as
Swihart velocities, $\widetilde{c}_{1,2}$. In the general case, Swihart
velocities are equal to\cite{LTgold}

\begin{equation}
\widetilde{c}_{1,2}=\frac{\sqrt{2}c_{01}c_{02}}{\sqrt{c_{01}^2+c_{02}^2\pm 
\sqrt{\left( c_{01}^2-c_{02}^2\right) ^2+4S^2c_{01}^2c_{02}^2}}},
\label{Eq11}
\end{equation}

where $c_{02}=c_{01}\sqrt{\frac{C_1\Lambda _1}{C_2\Lambda _2}}$ is the
Swihart velocity of the single JJ2.

The most crucial changes in the fluxon shape occur as the velocity
approaches the lowest Swihart velocity, $u\rightarrow \widetilde{c}_1$. Then
the first component is totally squeezed, $\lambda _1\rightarrow 0$, while
contraction of the second component remains marginal, see Eq.(8). In this
case the two components become clearly distinguishable: the $F_1$ component
transforms into a step-like function which changes from zero to 2$\pi $
within the distance $\lambda _1$ at the fluxon center, while outside the
central region the shape of the fluxon is defined by the $F_2$ component.
From Eq.(10) it follows that

\begin{equation}
\frac{sin\left( \varphi _1\right) }{sin\left( \varphi _2\right) }= 
{\kappa _2,\left| x\right| \gg \lambda _1; \atopwithdelims\{. -\kappa _1,x\rightarrow 0.}
\left( u\rightarrow \widetilde{c}_1\right) .  \label{Eq12}
\end{equation}
For $\frac{C_2\Lambda _2}{C_1\Lambda _1}=1$, and $u=\widetilde{c}_1$, the
parameters $\kappa _{1,2}$ are equal to:

\begin{equation}
\kappa _{1,2}=-\sqrt{\frac{\Lambda _1}{\Lambda _2}};\sqrt{\frac{\Lambda _2}{%
\Lambda _1}}\frac{J_{c2}}{J_{c1}}.  \label{Eq13}
\end{equation}
The parameters $\kappa _{1,2}$ determine the weight coefficients of the
components. From Eqs.(10,13) it follows that $F_1$ component dominates in
junction 1 for $J_{c2}/J_{c1}\ll 1$, and $F_2$ dominates for $%
J_{c2}/J_{c1}\gg 1$.

From the analysis above it is seen that the linearized coupled sine-Gordon
equation allows both the single component solutions, $F_{1,2}$, Eq.(6) and
the double component solution, Eq.(10). At this stage it is not clear which
of the solutions, Eqs.(6,10), should be realized in SJJ's, since all three
solutions have the same accuracy with respect to Eq.(1). In Refs.\cite
{Modes,Fluxon} it was shown that it is the double component solution Eq.(10)
which is realized in the static case. However, it was suggested that a
single component solution could be achieved at high fluxon velocities.
Indeed, as we will show below, in the dynamic case both single and double
component solutions can exist and even coexist, depending on parameters of
the stack and the fluxon velocity. What is important, however, is that these
are always the components $F_{1,2}$ described by Eqs.(6,7) which constitute
the fluxon.

III. FRICTIONLESS CASE

In this section we will consider unperturbed, $\alpha _i=J_b=0$,
frictionless fluxon motion. We analyze the pure solitonic fluxon motion for
various junction parameters, make general conclusions about transformation
of the fluxon shape in dynamics and compare it with analytical predictions.
The effect of finite damping and bias will be considered in section IV. For
numerical simulations of frictionless fluxon motion we used Eq.(3). The
numerical procedure was based on a finite difference method with successive
iterations of ODE in Eq.(3). The boundary conditions were such that the
total phase shift is equal to 2$\pi $ in the junction containing a fluxon
and zero in the other one. The fluxon will be placed in JJ1 if not stated
otherwise. We will consider five cases: A) a stack of identical junctions,
B) stack with different critical current densities, C) fluxon in the
junction with lower critical current density, D) fluxon in the junction with
the higher critical current density, E) a stack with difference both in
critical currents and electrodes. Finally in subsection III.F we derive and
verify conditions for observation of single and double component fluxon at $%
u\rightarrow \widetilde{c}_1$ for various parameters of SJJ's.

{\it A. Identical junctions: double component solution.}

In Fig.1 profiles of a) phase differences $\varphi _{1,2}$, b) the ratio $%
sin(\varphi _1)/sin(\varphi _2)$, and c) magnetic inductions $B_{1,2}$ of a
single fluxon in junction 1 are shown for a double stack consisting of
identical strongly coupled junctions and for different fluxon velocities, $u/%
\widetilde{c}_1$=0, 0.61, 0.92, 0.98, 0.998, 0.9999 (from left to right
curve). The curves were shifted for clarity along the {\it x}-axis.
Parameters of the stack are: $d_i=t_i=0.01\lambda _{J1}$, $\lambda
_{Si}=0.1\lambda _{J1}$, $S\simeq 0.5$, $C_1=C_2$. In Fig.1 a) dotted lines
show profiles obtained from the analytic double component solution Eq.(10).
Solid and dashed lines in Fig. 1 a,c) represent results of numerical
simulation for junctions 1 and 2, respectively. The data in Figs.1 b,c) are
obtained numerically. The magnetic induction is normalized to $H_0=\frac{%
\Phi _0}{\pi \lambda _{J1}\Lambda _1}$.

As the velocity approaches the lower Swihart velocity, $\widetilde{c}_1$,
the existence of the two fluxon components becomes clearly seen. For
identical junctions, as it follows from Eqs.(10,13), exactly one half of the
fluxon belongs to each component. The $F_1$ component transforms to a one-$%
\pi $ step. Outside the fluxon center the fluxon is defined entirely by the $%
F_2$ component, which is only marginally contracted. Moreover, from Fig.1 a)
it is seen that at $u\sim \widetilde{c}_1$ the phase differences in both
junctions are equal outside the fluxon center in agreement with analytical
prediction, Eq.(12). This is illustrated in Fig. 1 b), from which it is seen
that the ratio $sin(\varphi _1)/sin(\varphi _2)$ approaches unity as $%
u\rightarrow \widetilde{c}_1$. From Fig.1 a) it is seen that the approximate
analytic solution is in good agreement with numerical solution for all
fluxon velocities.

Another unusual feature of the moving fluxon in SJJ's is seen from Fig. 1
c). A dip in $B_2$ is developed with increasing fluxon velocity, leading to
inversion of the sign at high velocities. Such behavior was predicted
analytically in Ref.\cite{Modes}. Here we confirm the existence of this
phenomenon by numerical simulation.

{\it B. Nonidentical junctions: transformation from }$F_2${\it \ to }$F_1$ 
{\it component solution with decreasing }$J_{c2}/J_{c1}${\it :}

In Ref.\cite{Modes} it was shown that it is the double component solution
that is realized in the static case for arbitrary parameters of the stack.
From the numerical analysis of the dynamic case we have found that the
fluxon shape is well described by the double component solution up to
velocities very close to $\widetilde{c}_1$. This is illustrated by Fig.2, in
which the shape of the fluxon in junction 1 moving with the velocity close
to the lower Swihart velocity, $u=0.98\widetilde{c}_1$, is shown for
different critical currents, $J_{c2}/J_{c1}$=10; 2; 1; 0.5; 0.1, from left
to right curve. The rest of the parameters of the stack and the way of
presentation is the same as in Fig. 1a). From Fig.2 it is seen that the
fluxon shape is in a good agreement with the analytical double component
solution, Eq.(10), up to velocities very close to $\widetilde{c}_1$. Note,
that the agreement is even better at lower fluxon velocities. From Fig.2 it
can be seen that a gradual transformation of the fluxon shape from the
uncontracted $F_2$ component solution to the contracted $F_1$ component
solution takes place as $J_{c2}/J_{c1}$ decreases. In terms of the
analytical double component solution, Eq.(10), this is caused by a decrease
of the weight coefficient $\kappa _2$ , Eq.(13), of the $F_2$ component.

As the velocity approaches the lower Swihart velocity, $\widetilde{c}_1$,
the fluxon shape can remain the double-component, as shown in Fig.1, or a
transformation of the fluxon shape could take place. As we have found from
our numerical simulations the transformation and the final shape of the
fluxon at $u=\widetilde{c}_1$ strongly depend on parameters of the stack.

{\it C. Fluxon in a weaker junction: Uncontracted single component solution.}

In Fig. 3 the fluxon shape is shown for the case when the fluxon is placed
in the weaker junction, $J_{c2}/J_{c1}=2$, the rest of the parameters and
the way of presentation are the same as in Fig.1. As it is seen from Fig.3
at velocities up to 0.98$\widetilde{c}_1$ the shape of the fluxon is well
described by the double component solution, Eq.(10). At higher velocities
transformation to the single $F_2$ component solution, Eq. 6, with takes
place. This is illustrated in Fig. 3 b), from which it is seen that as $%
u\rightarrow \widetilde{c}_1$, $sin(\varphi _1)/sin(\varphi _2)\rightarrow
\kappa _2=2$ in the whole space region. From Fig. 3 c) it is seen that a dip
in $B_2$ is reduced with respect to that in Fig.1 c), due to absence of the
squeezed $F_1$ component.

{\it D. Fluxon in a stronger junction: two component solution.}

In Fig. 4 the fluxon shape is shown for the case when the fluxon is placed
in the stronger junction, $J_{c2}/J_{c1}=0.5$, the rest of the parameters
and the way of presentation are the same as in Fig. 1. It is seen that at
velocities up to 0.98$\widetilde{c}_1$ the shape of the fluxon is well
described by the double component solution, Eq.(10). At higher velocities
the fluxon still has contracted and uncontracted components $F_{1,2}$. The
existence of the two fluxon components is clearly seen from Fig. 4 b). As $%
u\rightarrow \widetilde{c}_1$, $sin(\varphi _1)/sin(\varphi _2)\rightarrow
\kappa _2=0.5$ outside the center of the fluxon and $sin(\varphi
_1)/sin(\varphi _2)\rightarrow -\kappa _1=1$ in the center, in agreement
with Eqs.(12,13). However, transformation of the fluxon shape with respect
to Eq.(10) takes place. From Fig.4 a) it is seen that in the left half-space
(i.e. from far left to the immediate left of the fluxon center) the phase
shift in junctions 1 and 2 approaches zero and $\pi $, respectively, and
belongs to the uncontracted $F_2$ component, as seen from Fig. 4b).
Therefore, there is a single $F_2$ component fluxon placed in junction 2
(the weaker junction). The situation in the left half space is then
analogous to that in Fig.3. This is shown in Fig.5, in which we replotted
the curves at $u=0.9999\widetilde{c}_1$ from Figs. 3a) and 4a). Solid and
dashed curves in Fig.5 represent $\varphi _1$ and $\varphi _2$,
respectively, for the fluxon in the stronger junction from Fig.4, while
dashed-dotted and dotted curves represent $\varphi _1$ and $\varphi _2$,
respectively, for the fluxon in the weaker junction from Fig.3. Note, that
in Fig.5 we normalized the $x$-axis to $\lambda _{J1}$ for the case $%
J_{c2}/J_{c1}=2$ and to $\lambda _{J2} $ for the case $J_{c2}/J_{c1}=0.5$,
since the $F_2$ component is then situated in junction 2. From Fig.5 it is
seen that after such rescaling $\varphi _{2,1}$ from Fig. 4a) merge with $%
\varphi _{1,2}$ from Fig. 3a) in the left half-space. In the central region
a step-like change of phase shift takes place in both junctions. In junction
1 the phase jumps on $+2\pi $, which means that there is a single $F_1$
component fluxon and in junction 2 the phase drops on $-2\pi $, representing
the single $F_1$ component antifluxon. The overall fluxon shape at $u=%
\widetilde{c}_1$ for the case when the fluxon is in the stronger junction,
Fig. 4, can be written as:

\begin{equation}
\left\{ 
\begin{array}{c}
\varphi _1=F_1+Image(\varphi _2) \\ 
\varphi _2=F_2-F_1
\end{array}
\right. ,  \label{Eq14}
\end{equation}

so that in junction 1 there is the contracted single component $F_1$ fluxon
plus an image in a form of a ripple from junction 2 and in junction 2 there
is an uncontracted fluxon $F_2$ - contracted antifluxon $F_1$ pair. The
total phase shifts in junction 1 and 2 are $2\pi $ and zero, respectively,
however, the phase shift in junction 1 increases nonmonotonously and has two
local maxima and minima, see Fig. 4 a). From Fig.4 c) it is seen that the
dip in $B_2$ in this case is even more pronounced than that for identical
SJJ's, Fig.1 c). This is due to the increase of the weight coefficient of
the squeezed $F_1$ component.

Fig. 6 shows profiles of the fluxon moving with velocity very close to the
lowest Swihart velocity, $u=0.9999\widetilde{c}_1$, for different critical
current densities $J_{c2}/J_{c1}=10\div 0.1$ increasing sequentially from
the left to the right curve. The rest of the stack parameters and the way of
presentation is the same as in Fig. 1. From Fig.6 it is seen how the shape
of the fluxon is changed with $J_{c2}/J_{c1}$. When the fluxon is placed in
the weaker junction, the fluxon shape at the lowest Swihart velocity is
described by the single $F_2$ component. For the case of Figs. 1-6, $%
C_2/C_1=\Lambda _2/\Lambda _1=1$, so that $\kappa _2=\frac{J_{c2}}{J_{c1}}$.
In Fig. 6 b) the dependence $sin(\varphi _1)/sin(\varphi _2)=\frac{J_{c2}}{%
J_{c1}}$ is clearly visible in the whole space region for $J_{c2}/J_{c1}>1$.
When the fluxon is placed in the stronger junction, it has two components, $%
F_{1,2}$. From Fig. 6 b) it is seen that for the case $J_{c2}/J_{c1}<1$, the
fluxon shape in the center is determined by the $F_1$ component, $%
sin(\varphi _1)/sin(\varphi _2)=-\kappa _1=1$, while outside the center the
shape is given by the $F_2$ component, $sin(\varphi _1)/sin(\varphi
_2)=\kappa _2=\frac{J_{c2}}{J_{c1}}$, confirming that the parameters of the
single component solutions are always those that predicted by Eqs.(6,7).

From Fig. 6 it is seen that the transition from a double to a single
component solution for $J_{c2}/J_{c1}>1$ is gradual. Outside the fluxon
center this transition is well described by a gradual increase of the weigh
coefficient of the $F_2$ component, Eq.(10,13). In the center the exact
shape of the fluxon can be obtained from the first integral, Eq.(5). For the
case $\frac{C_2\Lambda _2}{C_1\Lambda _1}=1$, at $u=\widetilde{c}_1$ the
first integral reduces to:

\begin{equation}
\frac S{1-S^2}\left( 1+\frac{J_{c1}\Lambda _1}{J_{c2}\Lambda _2}\frac{%
cos\left( \varphi _1\right) }{cos\left( \varphi _2\right) }\right) ^2\frac{%
\left( \varphi _{1\xi }^{\prime }\right) ^2}2=1-cos\left( \varphi _1\right) +%
\frac{J_{c2}}{J_{c1}}\left( 1-cos\left( \varphi _2\right) \right) ,
\label{Eq15}
\end{equation}

for the $F_2$ single component solution. From Eq.(15) it is seen that at the
fluxon center, $x=0$, the effective Josephson depth is equal to:

\begin{equation}
\lambda _{eff}(F_2)=\lambda _{J1}\left( 1-\frac{J_{c1}\Lambda _1}{%
J_{c2}\Lambda _2}\right) \sqrt{\frac S{1-S^2}},  \label{Eq16}
\end{equation}

so that $\lambda _{eff}$ gradually increases from zero to $\lambda _{J1}%
\sqrt{\frac S{1-S^2}},$as $\frac{J_{c2}\Lambda _2}{J_{c1}\Lambda _1}$
becomes larger than unity. The inequality

\begin{equation}
\frac{J_{c2}\Lambda _2}{J_{c1}\Lambda _1}>1,  \label{Eq17}
\end{equation}

is then a necessary (but as we'll show below not sufficient) condition for
the existence of the single component $F_2$ solution at $u=\widetilde{c}_1$,
for $\frac{C_2\Lambda _2}{C_1\Lambda _1}=1$, since then $\lambda _{eff}$
remains finite at $u=\widetilde{c}_1$.

On the other hand, the transition from a double component solution, Eq.(10),
to the solution, Eq.(14), for $J_{c2}/J_{c1}<1$ is sharp. However, the
closer is $J_{c2}/J_{c1}$ to unity, the closer must be the fluxon velocity
to $\widetilde{c}_1$ in order to observe this transformation, as it can be
seen from Figs. 4,6.

{\it E. Nonidentical electrodes: Bifurcations and more complicated two
component solutions.}

So far we have considered the case when only critical current densities of
JJ's in the stack were different. Another common type of the difference
between JJ's is the difference in the properties of electrodes. In Fig.7 the
fluxon shape is shown for the case, $\Lambda _2\simeq 2.5\Lambda _1$.
Physically this means that the third electrode has either larger London
penetration depth, $\lambda _{S3}=2\lambda _{J1,2}$, or it is thinner than
the rest of the electrodes, $d_{1,2}=4d_3$, see definitions in sec.II. The
rest of the parameters are $J_{c2}/J_{c1}=0.5$, $d_{1,2}=t_i=0.01\lambda
_{J1}$, $\lambda _{S1,2}=0.1\lambda _{J1}$, $S\simeq 0.31$, $\frac{%
C_2\Lambda _2}{C_1\Lambda _1}=1$. The way of presentation is the same as in
Fig. 1. As it is seen from Fig.7, at velocities up to 0.98$\widetilde{c}_1$
the shape of the fluxon is well described by the analytic double component
solution, Eq.(10). From Fig. 7b) it is seen that outside the fluxon center
the phase distribution is determined by the $F_2$ component with $\kappa
_2\simeq 0.79$ given by Eq.(13). However, at $u\sim 0.998\widetilde{c}_1$
the system exhibit bifurcations and a sudden switching to the solution given
by Eq.(14) occurs. At slightly larger velocity another bifurcation takes
place resulting in switching to yet another solution.

The switching between the solutions is hysteretic. If we start reducing the
fluxon velocity, the switching back to the double component solution takes
place at somewhat lower velocity. Therefore there is a certain region of
fluxon velocities for which those solutions coexist. Such situation is
illustrated in Figs.8 a-c), in which phase distributions $\varphi _{1,2}$
for three possible solutions are shown for $u\simeq 0.998\widetilde{c}_1$
and for the same stack as in Fig.7. As usual, solid and dashed curves in
Fig. 8 represent $\varphi _1$ and $\varphi _2$, respectively, obtained from
numerical simulations.

Fig 8 a) shows the solution coming from low velocities. Dashed-dotted and
dotted curves in Fig. 8 a) represent $\varphi _1$ and $\varphi _2$,
respectively, given by the approximate double component solution, Eq.(10),
showing good agreement with numerical solution.

With increasing velocity, switching to the solution of the type given by
Eq.(14) takes place. This solution is shown in Fig. 8 b). This solution
exist only in a very narrow velocity interval and with further increase of
velocity a switching to a new solution occurs. The new solution is shown in
Fig. 8 c). It is characterized by a large phase variation in the second
junction. Such solution exists up to $\widetilde{c}_1$.

Looking at the fluxon shape at $u=0.9999\widetilde{c}_1$ from Fig.7 we see
that it consists of three parts: (i) at the fluxon center, $x=0$, in
junction 1 there is a $2\pi $ step, which as we will show in a moment,
belongs to the pure $F_1$ component and in junction 2 there is a $4\pi $ (!)
drop belonging to two flux quantum, $-2F_1$, antifluxon. (ii) At $%
x=x_0\simeq 0.7\lambda _{J1}$ and (iii) $x=-x_0$ in junction 2 there is a $%
2\pi $ increase and in junction 1 there is a ripple like phase shift. As it
can be seen from Figs. 7 a,c) the features at $x=\pm x_0$ contain both
contracted and uncontracted parts and are in fact given by the double
component solutions, Eq.(10). This is illustrated in Fig.9 a) in which solid
and dashed lines show phase distributions $\varphi _1$ and $\varphi _2$,
respectively, from Fig. 7 a) for $u=0.9999\widetilde{c}_1$ and dashed-dotted
and dotted curves represent $\varphi _1$ and $\varphi _2$, respectively,
given by the analytic double component solution, Eq.(10), shifted by $-x_0$
along the $x$- axis. Obviously, the features at $x=\pm x_0$ correspond to
the double component soliton placed in junction 2. Due to spatial separation
between the contracted centrum of the fluxon and the double component
features at $x=\pm x_0$, we can analyze the shape of the central contracted
part. From Fig. 7 c) it is seen that the magnetic induction in junction 1 at 
$x=0$, $B_1(0)$, increases sharply as the velocity approaches $\widetilde{c}%
_1$. If the phase distribution in the central region is given by the $F_1$
component, then from Eq. (2) it follows that $B_1(0)\sim F_1^{\prime
}(0)\sim \lambda _1^{-1}\sim \left( 1-\frac{u^2}{\widetilde{c}_1^2}\right)
^{-1/2}$, representing Lorentz contraction of $F_1$ component at the lowest
Swihart velocity, $\widetilde{c}_1$. Fig. 9 b) shows the inverse value of $%
B_1(0)$ versus the Lorentz factor $\left( 1-\frac{u^2}{\widetilde{c}_1^2}%
\right) ^{1/2}$, for the high velocity solution from Fig.(7). Dots represent
the numerically obtained values, solid line is the apparent linear fit.
Lorentz contraction of the central region at the lowest Swihart velocity, $%
\widetilde{c}_1$, is clearly seen from Fig. 9 b), therefore confirming that
the central region is given by the pure $F_1$ component. The overall shape
of the soliton in this case can be described as:

\begin{equation}
\left\{ 
\begin{array}{c}
\varphi _1=F_1+Image(\varphi _2) \\ 
\varphi _2=F_{1+2}(x-x_0)-2F_1+F_{1+2}(x+x_0)
\end{array}
\right. ,  \label{Eq18}
\end{equation}

where $F_{1+2}$ is the double component solution, given by Eq.(10). From
Eq.(18) it is seen that the overall phase shift is $2\pi $ in junction 1 and
zero in junction 2, however the phase growth in junction 1 is nonmonotonous.

{\it F. Conditions for the existence of single and two component solutions.}

For practical applications of SJJ's in flux-flow oscillators, the shape of
the fluxon at the highest propagation velocity is crucial. It is then
important to know how the shape of the fluxon at $u=\widetilde{c}_1$ depends
on the parameters of the stack. Eq.(17) formulates the necessary condition
for the existence of uncontracted single component $F_2$ solution at $u=%
\widetilde{c}_1$, for $\frac{C_2\Lambda _2}{C_1\Lambda _1}=1$, since then $%
\lambda _{eff}$ remains finite at $u=\widetilde{c}_1$. However this
condition is not sufficient. Indeed, for the case of Figs.7,8, $\frac{%
J_{c2}\Lambda _2}{J_{c1}\Lambda _1}\simeq 1.25$, i.e. Eq.(17) is satisfied.
However, we obviously do not observe the single component $F_2$ solution at $%
u=\widetilde{c}_1$, but rather a switching to more complicated two component
solutions, Eqs.(14,18), takes place. From Fig. 8 a) it is seen that the
switching takes place when maximum value of $\varphi _2$ approaches $\pi /2$%
. If the solution were to stay at the special single component $F_2$
solution, then according to Eq.(6), the maximum value of $sin(\varphi _2)$
would be equal to $1/\kappa _2$. This gives us an additional condition for
the observation of the single component $F_2$ solution:

\begin{equation}
\kappa _2>1,  \label{Eq19}
\end{equation}

In Fig.10, regions of the existence of the two component (shaded area) and
the single component $F_2$ solutions at $u=\widetilde{c}_1$ are shown for $%
\frac{C_2\Lambda _2}{C_1\Lambda _1}=1$. Numbers in Fig. 10 show the number
of components obtained numerically. Solid and dashed lines represent the
conditions Eq.(19) and Eq.(17), respectively. Arrows indicate the cases
considered in Figs. 1,3,4,7. It is seen that the single component solution
exists when both conditions, Eqs.(17,19), are satisfied.

IV. IMPLEMENTATION FOR EXPERIMENTAL SITUATION

In the previous section we have studied the unperturbed fluxon motion, $%
\alpha _i=0,J_b=0$. This is an idealized case. In real experimental
situation $\alpha _i\neq 0$ and $J_b\neq 0$. To quantitatively study the
influence of damping and bias on the fluxon shape and IVC's, we performed
numerical simulation of the coupled sine-Gordon equation, Eq.(1), with the
dissipation and bias terms. Here we used two different approaches: (i)
Considering solitonic-type fluxon motion, $\varphi _i=\varphi _i(\xi )$, we
derive the system of one-dimensional ordinary differential equations (ODE),
similar to Eq.(3), but with dissipation terms, $\alpha _i\neq 0$, and bias
terms, and used the same numerical procedure to solve it. (ii)
Alternatively, we directly integrated the system of partial differential
equations (PDE), Eq. (1) using explicit finite difference method. Both
approaches have certain advantages and disadvantages. Using ODE, it is
possible to calculate fluxon shape for arbitrary small damping, while PDE
require relatively large damping coefficient. In addition, ODE in comparison
to PDE, does not have problem with accumulation of error and does not
require a long relaxation and averaging times. For solitonic fluxon motion
both approaches give identical results. On the other hand, ODE are
restricted to the study of solitonic motion, while the PDE allow more
complicated solutions. For simulation of PDE, periodic boundary conditions
were used, which correspond to fluxon motion in annular SJJ's with $L=10\div
100\lambda _{J1}$.

A{\it . Effect of damping and current-voltage characteristics.}

First of all fluxon shape will affect the shape of the flux-flow
current-voltage characteristics (IVC's) caused by fluxon motion in the stack%
\cite{ZFStep}. The shape of flux-flow IVC is determined by a balance between
the input power to the system from the current bias source and the power
dissipated due to a finite damping, see e.g. Ref.\cite{Scott}. In a single
JJ's as the fluxon velocity approaches the Swihart velocity, Lorentz
contraction takes place and the fluxon energy increases sharply. The fluxon
velocity asymptotically approaches the Swihart velocity with increasing
current. In the IVC this result in the appearance of an almost vertical step
at the velocity matching condition\cite{Scott}. Since the existence of this
step is closely related to the Lorentz contraction of the fluxon, we expect
that the step at the velocity matching condition should also exist in SJJ's
with the characteristic velocity equal to the lower Swihart velocity, $%
\widetilde{c}_1$, whenever the fluxon contains the Lorentz contracted part $%
F_1$. On the other hand, when a pure $F_2$ component solution takes place,
the fluxon will reach $u=\widetilde{c}_1$ at a finite current and flux-flow
IVC should have a finite slope at $u=\widetilde{c}_1$. For the case when the
transformation of the fluxon shape given by Eq.(18) (and probably by Eq.(14)
as well ) takes place, this would result in a premature switching from the
flux-flow branch and, possibly, in the existence of an extra metastable
flux-flow branch in the IVC with the same limiting velocity, $u=\widetilde{c}%
_1$, but with larger dissipation.

The average DC voltage in JJ 1 is $V_1/V_{01}=u/c_{01}$, where $V_{01}=\frac{%
\hbar \pi c_{01}}{2eL}$, and in JJ 2 is zero. Therefore, we now plot
current-velocity characteristics to represent IVC's.

In Fig. 11, the single fluxon IVC's are shown for double stacks with equal
damping coefficients, $\alpha _{1,2}=0.05$, and for $J_{c2}/J_{c1}=1$,
(solid diamonds); $J_{c2}/J_{c1}=0.5$, (open circles); $J_{c2}/J_{c1}=2$,
(open squares). The rest of parameters are the same as in Fig. 1. In Fig.
11, symbols represent solutions obtained from ODE and subsequent solid lines
show solution of the full PDE, Eq. (1), dashed gray line shows the IVC of an
uncoupled single junction 1 and dotted line indicates the position of the
lower Swihart velocity, $\widetilde{c}_1\simeq 0.817c_{01}$. Insets in Fig.
11 show spatial distribution of $sin(\varphi _1)$ (solid lines), and $%
sin(\varphi _2)$ (dashed lines), for maximum propagation velocities, for
which ODE based numerical procedure converged: $u_{max}(J_{c2}/J_{c1}=1)%
\simeq 0.999\widetilde{c}_1$, $u_{max}(J_{c2}/J_{c1}=0.5)\simeq 0.97%
\widetilde{c}_1$, $u_{max}(J_{c2}/J_{c1}=2)\simeq 0.99\widetilde{c}_1$.
Solid lines in Fig.11 show that PDE allow solutions propagating with even
larger velocity, however those solutions are not of solitonic type, and will
be discussed in the next section. From Fig. 11 it is seen that the IVC's for 
$J_{c2}/J_{c1}=1$ and $0.5$, exhibit velocity matching behavior at $%
u\rightarrow \widetilde{c}_1$. On the other hand, for $J_{c2}/J_{c1}=2$, no
velocity matching behavior is observed, and the velocity reaches $\widetilde{%
c}_1$ at a finite current. Such behavior is in agreement with the absence of
contracted $F_1$ component, as discussed above. From the insets in Fig. 11
it is seen, that the fluxon shape for the case of small damping does not
differ much from the frictionless case, considered in the previous section.
On the other hand, damping reduces the stability of the fluxon state at high
propagation velocities, so that the maximum fluxon velocity for pure
solitonic motion becomes less than $\widetilde{c}_1$. This is due to the
finite bias current in the stack, which result in asymmetry of phase
distribution in the stack. Such asymmetry is clearly seen from inset in Fig.
11 for $J_{c2}/J_{c1}=0.5$. From Fig. 11 it is seen, that reduction of
stability in the flux-flow state depends on parameters of the stack. E.g.
for the case of double stack with nonidentical junctions, flux-flow state is
more stable for fluxon in the weaker junction, $J_{c2}/J_{c1}=2$, than for
fluxon in the stronger junction, $J_{c2}/J_{c1}=0.5$. For the case $%
J_{c2}/J_{c1}=0.5$, switching of the second junction to the quasiparticle
branch occurs first at $I/I_{c1}\simeq 0.22$.

In real life of course junctions in the stack are not prenumerated and if
junctions are not identical, the stable state will correspond to a fluxon
placed in the weaker junction. Therefore whenever properties of junctions in
the stack are considerably different, the stable dynamic state at $u=%
\widetilde{c}_1$ would correspond to the existence of uncontracted $F_2$
component soliton in the weaker junction. Experimentally, steps at the
velocity matching condition were observed for low-$T_c$ SJJ's\cite
{Sakai,Monaco}, on the other hand, for high-$T_c$ intrinsic SJJ's the
flux-flow IVC's always have a finite slope\cite{Lee}. We note, that for
stacks with large number of junctions with thin electrodes, $d\ll \lambda _s$%
, it is not necessary to have a variation in $J_{ci}$ to make the junctions
different. In this case the middle junctions have a lower critical field, $%
H_{c1}$, approximately half of that compared to the outmost junctions due to
the fact that fluxon in the outmost junctions carries only half a flux
quantum. In a double stack, considered here, a corresponding thing happens
when the junctions have different electrode thicknesses. The junction with
thicker electrodes may have lower $H_{c1}$ even if $J_c$ of this junction is
larger, see Ref. \cite{Modes}. In a sense, the criterion for transition from
'weak' to 'strong' junction is given by Eqs. (17,19).

Although the stable state corresponds to the case when fluxon is placed in
the weaker junction, the situation when the fluxon is placed in the stronger
junction can also be achieved in experiment as shown in Ref.\cite{Monaco}.
Obviously such state can be achieved in annular SJJ's. In this case the
fluxon can eventually be introduced in the stronger junction and if so it
will stay there and can be accelerated by the bias current.

{\it C. Velocity above }$\widetilde{c}_1${\it : Cherenkov radiation. }

So far we have considered the case $u\leq \widetilde{c}_1$. For $u>%
\widetilde{c}_1$, coefficients before the second derivative of phase in
Eq.(3) become negative. The equation can still be written in a sine-Gordon
type form if we substitute $\varphi (\xi )=\pi +\phi (\xi )$. Indeed, a 2$%
\pi $ soliton like solution for $\phi (\xi )$, moving with $u>\widetilde{c}%
_1 $, does exist. However the energy of this state is decreasing with
increasing velocity and therefore such state would correspond to unstable
IVC branch with negative resistance.

In Ref.\cite{Modes} it was suggested that for $u>\widetilde{c}_1$ the fluxon
is a combination of a soliton with Josephson plasma waves. From Eq. (8) it
is seen that $\lambda _1$ becomes imaginary for $u>\widetilde{c}_1$ and the $%
F_1$ component transforms into a travelling Josephson plasma wave. The
fluxon solution is then given by the $F_2$ component soliton accompanied by
Josephson plasma waves from the degenerate $F_1$ component. Recently, the
existence of such type of solution was shown by numerical simulation in Ref.%
\cite{Cheren} and was interpreted as Cherenkov radiation in SJJ's, when
fluxon velocity exceeds the phase velocity of electromagnetic waves. This
solution is not of soliton type and can not be obtained from ODE. Therefore
solution of full PDE, Eq.(1), is required. From Fig. 11 it is seen that for
the case $J_{c2}/J_{c1}=2$, PDE allow solutions propagating with $u>%
\widetilde{c}_1$.

In Fig.12 results of numerical simulations of PDE, Eq.(1), for the case $%
J_{c2}/J_{c1}=2$, $u>\widetilde{c}_1$ are shown. Parameters of SJJ's are the
same as in Fig.11. Insets show spatial distributions of $sin(\varphi _1)$
(solid lines) and $sin(\varphi _2)$ (dashed lines) for a) $u/\widetilde{c}%
_1\simeq 1.015$, $I/I_{c1}=0.15$ and b) $u/\widetilde{c}_1\simeq 1.155$, $%
I/I_{c1}=0.5$, respectively. Simulations were done for annular SJJ's with $%
L=100\lambda _{J1}$. From our simulations we observe that fluxon shape
changes gradually as $u$ exceeds $\widetilde{c}_1$. Therefore, there are no
peculiarities at $u=\widetilde{c}_1$ in the IVC, see solid line in Fig.11
for $J_{c2}/J_{c1}=2$. Indeed, from inset a) in Fig. 12 it is seen that for $%
u$ slightly above $\widetilde{c}_1$, fluxon shape in the left half-space is
similar to that at $u<\widetilde{c}_1$, see the bottom inset in Fig.11.
However, small oscillations appear behind the fluxon (fluxon is propagating
from right to left). As the velocity increases, both amplitude and
wavelength of the oscillations increase, as illustrated in inset b) in Fig.
12. To clarify the physical origin of the oscillations, in Fig. 12 we have
plotted the average wavelength of oscillations (circles) as a function of
the absolute value of the Lorentz factor, $\sqrt{(u/\widetilde{c}_1)^2-1}$.
The solid line in Fig. 12 shows the absolute value, $2\pi \left| \lambda
_1\right| $, given by Eq.(8) and describing small amplitude plasma waves
from the degenerate $F_1$ component\cite{Modes} (the factor $2\pi $ is due
to different definition of the wavelength and the penetration depth).
Excellent agreement between the wavelength of Cherenkov radiation and
Josephson plasma wavelength from the degenerate $F_1$ component is observed
without any fitting, thus confirming the idea of Ref. \cite{Modes} that
Cherenkov radiation is due to plasma wave generation from the degenerate $F_1
$ component. We would like to note, that $\left| \lambda _1\right| $ is not
linear as a function of the Lorentz factor, although deviations from the
linear dependence are small. For high propagation velocities, an uncertainty
appears in determination of $\lambda $. This is caused by the increase of
the amplitude of oscillations, see inset b) in Fig. 12. Here oscillations
are not exactly monochromatic but the wavelength slightly increase with the
amplitude. The uncertainty in determination of $\lambda $ at high velocities
is shown by error bars in Fig.12.

CONCLUSIONS

In conclusion, we have shown that the shape of a single fluxon in double
stacked Josephson junctions can be described by existence of two components
given by Eqs. (6,7) and with characteristic lengths and velocities given by
Eqs.(8,9). At velocities up to $u\sim 0.98\widetilde{c}_1$, the soliton
shape is well described by the approximate double component solution,
Eq.(10), for all studied junction parameters, as illustrated in Figs.1-4,7.
In the very vicinity of the lower Swihart velocity the fluxon shape may
undergo radical transformations. The final shape of the fluxon at $u=%
\widetilde{c}_1$ strongly depends on parameters of the stack. From our
numerical simulations we have found that the fluxon may remain double
component, as shown in Fig.1, transform to a pure $F_2$ component solution,
see Fig.3, or be a more complicated combination of $F_1$ and $F_2$
components, see Eqs.(14,18) and Figs. 4,7,8. Those more complicated
solutions do not, strictly speaking, represent the single fluxon state, but
are combinations with fluxon-antifluxon pairs. However, even in this case
these are always the components $F_{1,2}$ described by Eqs.(6,7) that
constitute the solution. This implies that the components $F_{1,2}$ are real
and may live their own life and appear in different combinations. Conditions
for observing the single and the two-component solutions at $u=\widetilde{c}%
_1$ were formulated and verified. We have shown that as the velocity
approaches $\widetilde{c}_1$, the phase shift may become nonmonotonous.

A prominent feature of a soliton moving at the velocity close to $\widetilde{%
c}_1$ in SJJ's is the possible inversion of magnetic field $B_2(0)$, see
Figs.1,3,4 c). Such behavior was predicted analytically in Ref.\cite{Modes}.
Here we confirmed the existence of this phenomenon by numerical simulation.
The inversion of magnetic field in junction 2 may lead to attractive fluxon
interaction for fluxons in different junctions. Then the so-called
'in-phase' or 'bunched' state\cite{SBP} with fluxons one on the top of the
other in adjacent junctions may become favorable at high enough fluxon
velocity. In experiment this would result in appearance of the extra
flux-flow branch in the current-voltage characteristics with higher voltage,
as shown by numerical simulations \cite{Petrag} and observed in experiment
on low-$T_c$ SJJ's\cite{Sakai}. In Ref. \cite{GrJen} it was shown that the
bunched state can be stable at $u>\widetilde{c}_1$, however, no mechanism
for overcoming the mutual fluxon repulsion and transformation into the
bunched state was suggested. The existence of the field inversion might be a
criterion for the appearance of the bunched state in SJJ's. As we have
shown, the sign inversion and a dip in $B_2(0)$ disappears when junction 2
becomes considerably stronger than junction 1 and transformation of the
fluxon shape to a single $F_2$ component solution takes place, see Fig. 3 c).

The shape of the flux-flow IVC's was analyzed for various parameters of
SJJ's and it was shown that velocity matching behavior at $u=\widetilde{c}_1$
is observed when fluxon contains the contracted $F_1$ component. Finally,
Cherenkov radiation at $u>\widetilde{c}_1$ was shown to be due to generation
of plasma waves from the degenerate $F_1$ component in agreement with the
prediction of Ref. \cite{Modes}. Analytic expression for the wavelength of
the Cherenkov radiation is derived.

The work was supported by Swedish Superconductivity Consortium and in part
by the Russian Foundation for Basic Research under Grant No. 96-02-19319.

\begin{figure}[tbp]
\caption{ Profiles of a) phase differences $\varphi _{1,2}$, b) the ratio $%
sin(\varphi _1)/sin(\varphi _2)$, and c) magnetic inductions $B_{1,2}$ of a
single fluxon are shown for double SJJ's consisting of identical strongly
coupled junctions and for different fluxon velocities, $u/\widetilde{c}_1$%
=0, 0.61, 0.92, 0.98, 0.998, 0.9999 (from left to right curve). In Fig.1 a)
dotted lines show profiles obtained from the analytic double component
solution Eq.(10). The rest of the curves represent results of numerical
simulation. It is seen that the fluxon shape in this case is well described
by Eq.(10) and consist of contracted and uncontracted components. The sign
inversion of $B_2(0)$ at $u\simeq \widetilde{c}_1$ is clearly seen.}
\label{Fig.1}
\end{figure}

\begin{figure}[tbp]
\caption{ Fluxon shape at $u=0.98\widetilde{c}_1$, for different critical
currents, $J_{c2}/J_{c1}$=10; 2; 1; 0.5; 0.1, from left to right curve. The
rest of the parameters of the stack and the way of presentation is the same
as in Fig. 1a). It is seen that fluxon shape is in a good agreement with the
analytical double component solution, Eq.(10), up to velocities very close
to $\widetilde{c}_1$. A gradual transformation of the fluxon shape from the
uncontracted $F_2$ component solution to the contracted $F_1$ component
solution can be seen as $J_{c2}/J_{c1}$ decreases. }
\label{Fig.2}
\end{figure}

\begin{figure}[tbp]
\caption{ Fluxon shape for the case when the fluxon is placed in the weaker
junction, $J_{c2}/J_{c1}=2$, the rest of the parameters and the way of
presentation are the same as in Fig.1. At velocities up to 0.98$\widetilde{c}%
_1$ the shape of the fluxon is well described by the double component
solution, Eq.(10). At higher velocities transformation to the single $F_2$
component solution, Eq. (6), with $\kappa _2=2$, takes place, as seen from
Fig. 3 b). }
\label{Fig.3}
\end{figure}

\begin{figure}[tbp]
\caption{ Fluxon shape for the case when the fluxon is placed in the
stronger junction, $J_{c2}/J_{c1}=0.5$, the rest of the parameters and the
way of presentation are the same as in Fig. 1. It is seen that at velocities
up to 0.98$\widetilde{c}_1$ the shape of the fluxon is well described by the
double component solution, Eq.(10). At higher velocities the fluxon still
has two contracted and uncontracted components $F_{1,2}$. The existence of
the two fluxon components is clearly seen from Fig. 4 b). As $u\rightarrow 
\widetilde{c}_1$, $sin(\varphi _1)/sin(\varphi _2)\rightarrow \kappa _2=0.5$
outside the center of the fluxon and $sin(\varphi _1)/sin(\varphi
_2)\rightarrow -\kappa _1=1$ in the center. However, transformation of the
fluxon shape to that given by Eq.(14) takes place. }
\label{Fig.4}
\end{figure}

\begin{figure}[tbp]
\caption{ Phase distributions at $u=0.9999\widetilde{c}_1$ from Figs. 3a)
and 4a). Solid and dashed curves represent $\varphi _1$ and $\varphi _2$,
respectively, for the fluxon in the stronger junction $J_{c2}/J_{c1}=0.5$,
while dashed-dotted and dotted curves represent $\varphi _1$ and $\varphi _2$%
, respectively, for the fluxon in the weaker junction $J_{c2}/J_{c1}=2$. It
is seen that $\varphi _{2,1}$ from Fig. 4a) merge with $\varphi _{1,2}$ from
Fig. 3a) in the left half-space. Showing that the overall fluxon shape at $u=%
\widetilde{c}_1$ for the case when the fluxon is in the stronger junction,
Fig. 4, is given by Eq. (14). }
\label{Fig.5}
\end{figure}

\begin{figure}[tbp]
\caption{ Profiles of the fluxon moving with velocity very close to the
lowest Swihart velocity, $u=0.9999\widetilde{c}_1$, for different critical
current densities $J_{c2}/J_{c1}=10\div 0.1$ increasing sequentially from
the left to the right curve. The rest of the stack parameters and the way of
presentation of data is the same as in Fig. 1. When the fluxon is placed in
the weaker junction, the fluxon shape at the lowest Swihart velocity is
described by the single $F_2$ component. In Fig. 6 b) the dependence $%
sin(\varphi _1)/sin(\varphi _2)=\kappa _2=\frac{J_{c2}}{J_{c1}}$ is clearly
visible in the whole space region for $J_{c2}/J_{c1}>1$. When the fluxon is
placed in the stronger junction, it has two components, $F_{1,2}$. From Fig.
6 b) it is seen that for the case $J_{c2}/J_{c1}<1$, the fluxon shape in the
center is determined by the $F_1$ component, $sin(\varphi _1)/sin(\varphi
_2)=-\kappa _1=1$, while outside the center the shape is given by the $F_2$
component, $sin(\varphi _1)/sin(\varphi _2)=\kappa _2=\frac{J_{c2}}{J_{c1}}$%
. }
\label{Fig.6}
\end{figure}

\begin{figure}[tbp]
\caption{ Fluxon shape for the case of nonidentical electrodes, $\Lambda
_2\simeq 2.5\Lambda _1$ and $J_{c2}/J_{c1}=0.5$. The rest of the parameters
and the way of presentation is the same as in Fig. 1. It is seen that at
velocities up to 0.98$\widetilde{c}_1$ the shape of the fluxon is well
described by the double component solution, Eq.(10). From Fig. 7b) it is
seen that outside the fluxon center the phase distribution is determined by
the $F_2$ component with $\kappa _2\simeq 0.79$ given by Eq.(13). However,
at $u\sim 0.998\widetilde{c}_1$ a sudden switching to the other type of
solutions occur.}
\label{Fig.7}
\end{figure}

\begin{figure}[tbp]
\caption{ Illustration of coexistence of three types of solution for the
stack from Fig.7 for $u\simeq 0.998\widetilde{c}_1$. a)$\varphi _{1,2}$ for
the solution coming from low velocities, b) and c) show solutions given by
Eq.(14)and Eq.(18), respectively, coming from higher velocities. Solid and
dashed curves represent $\varphi _1$ and $\varphi _2$, respectively,
obtained from numerical simulations, while dashed-dotted and dotted curves
in Fig. 8 a) represent $\varphi _1$ and $\varphi _2$, respectively, given by
the approximate double component solution, Eq.(10).}
\label{Fig.8}
\end{figure}

\begin{figure}[tbp]
\caption{ a) Soliton shape at $u=0.9999\widetilde{c}_1$ from Fig.7. Solid
and dashed lines represent $\varphi _{1,2}$ obtained from numerical
simulation, while dashed-dotted and dotted lines represent the analytic
double component solution shifted by $x=-x_0$ and illustrating that the
features at $x=\pm x_0 $ correspond to the double component soliton in JJ 2.
b) Inverted values of magnetic induction in the center of junction 1 versus
the Lorentz factor. Symbols represent numerical simulations and solid line
is an apparent linear fit. Clear Lorentz contraction of the central region
at $u=\widetilde{c}_1$ is seen, showing that it belongs to a pure $F_1$
component. From Figs. 9 a,b) it is seen that the fluxon shape is given by
Eq.(18).}
\label{Fig.9}
\end{figure}

\begin{figure}[tbp]
\caption{Regions of existence of the two component (shaded area) and the
single component solutions at $u=\widetilde{c}_1$ for $\frac{C_2\Lambda _2}{%
C_1\Lambda _1}=1$. Numbers show the number of components obtained
numerically. Solid and dashed lines represent the conditions Eq.(19) and
Eq.(17), respectively. It is seen that the single component solution exists
when both conditions are satisfied. Arrows indicate the cases considered in
Figs. 1,3,4,7.}
\label{Fig.10}
\end{figure}

\begin{figure}[tbp]
\caption{Single fluxon IVC's are shown for double stack with equal damping, $%
\alpha _{1,2}=0.05$, and for $J_{c2}/J_{c1}=$0.5, 1, and 2. The rest of
parameters are the same as in Fig.1. Symbols and solid lines represent
results obtained by solving ODE, Eq.(3) and PDE, Eq.(1), respectively.
Dashed gray line shows the IVC of a single JJ 1 and dotted line shows the
position of the lower Swihart velocity, $\widetilde{c}_1$. Insets show
spatial distribution of $sin(\varphi _{1,2})$, for maximum propagation
velocities, }
\label{Fig.11}
\end{figure}

\begin{figure}[tbp]
\caption{The average wavelength of Cherenkov oscillations (circles) is shown
as a function of the absolute value of the Lorentz factor. The solid line
represents wavelength $2\pi \left| \lambda _1\right| $, for plasma waves
from the degenerate $F_1$ component, given by Eq.(8). Insets show spatial
distributions of $sin(\varphi _1)$ (solid lines) and $sin(\varphi _2)$
(dashed lines) for a) $u/\widetilde{c}_1\simeq 1.015$, $I/I_{c1}=0.15$ and
b) $u/\widetilde{c}_1\simeq 1.155$, $I/I_{c1}=0.5$, respectively. Arrows
indicate points in the main graph for which profiles were obtained. From
Fig.12 it is seen that Cherenkov radiation at $u>\widetilde{c}_1$ is caused
by plasma waves from the degenerate $F_1$ component.}
\label{Fig.12}
\end{figure}


\begin{references}
\bibitem{asin}  In Eq.(18) from Ref. \cite{Modes} the factor $1/\kappa
_{1,2} $ was taken out from $arcsin$, which is valid for $\varphi _2\ll 1$.
\bibitem{ZFStep}  Strictly speaking, we are talking about fluxon steps in
annular SJJ's and zero-field steps.
\end{references}
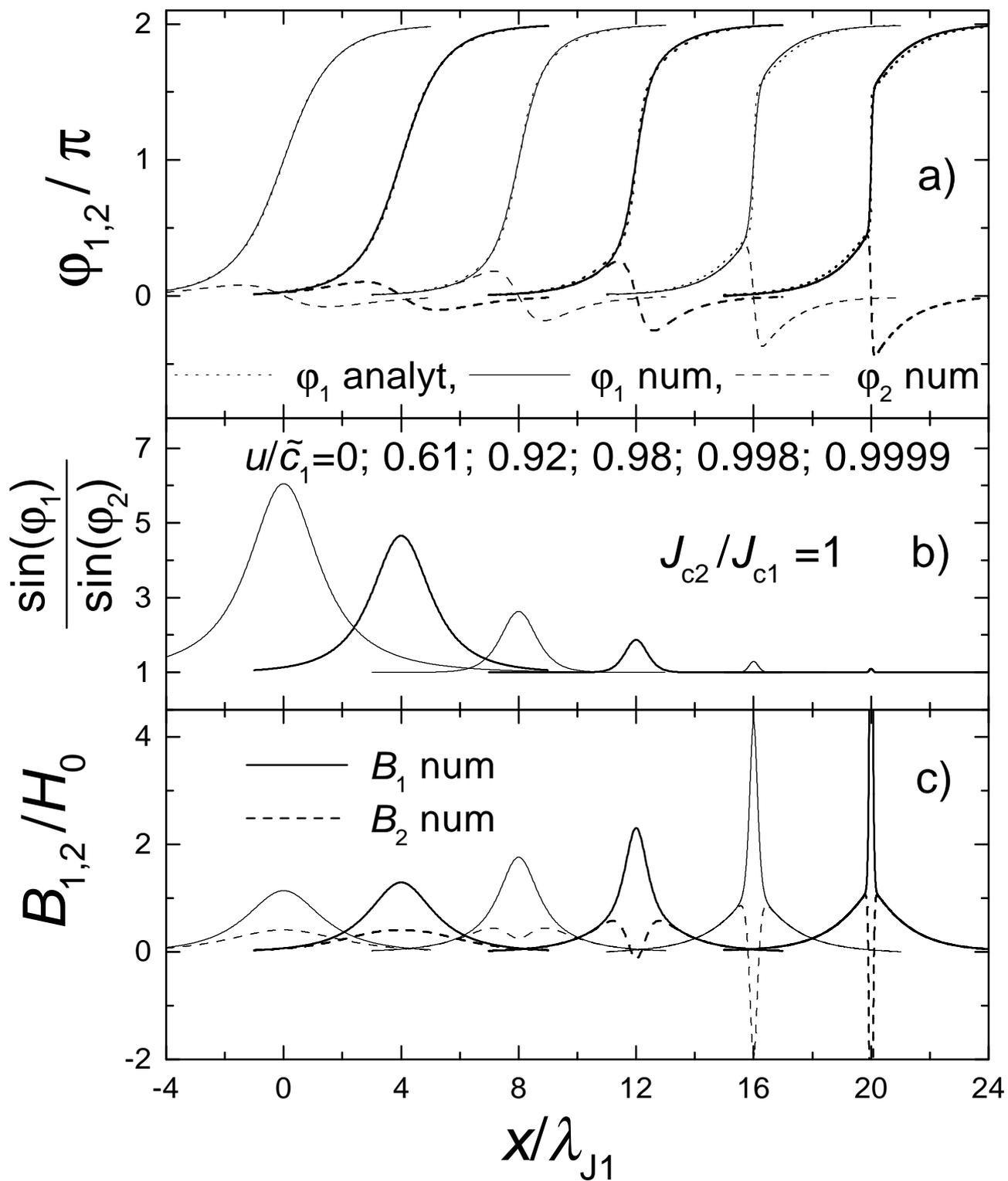
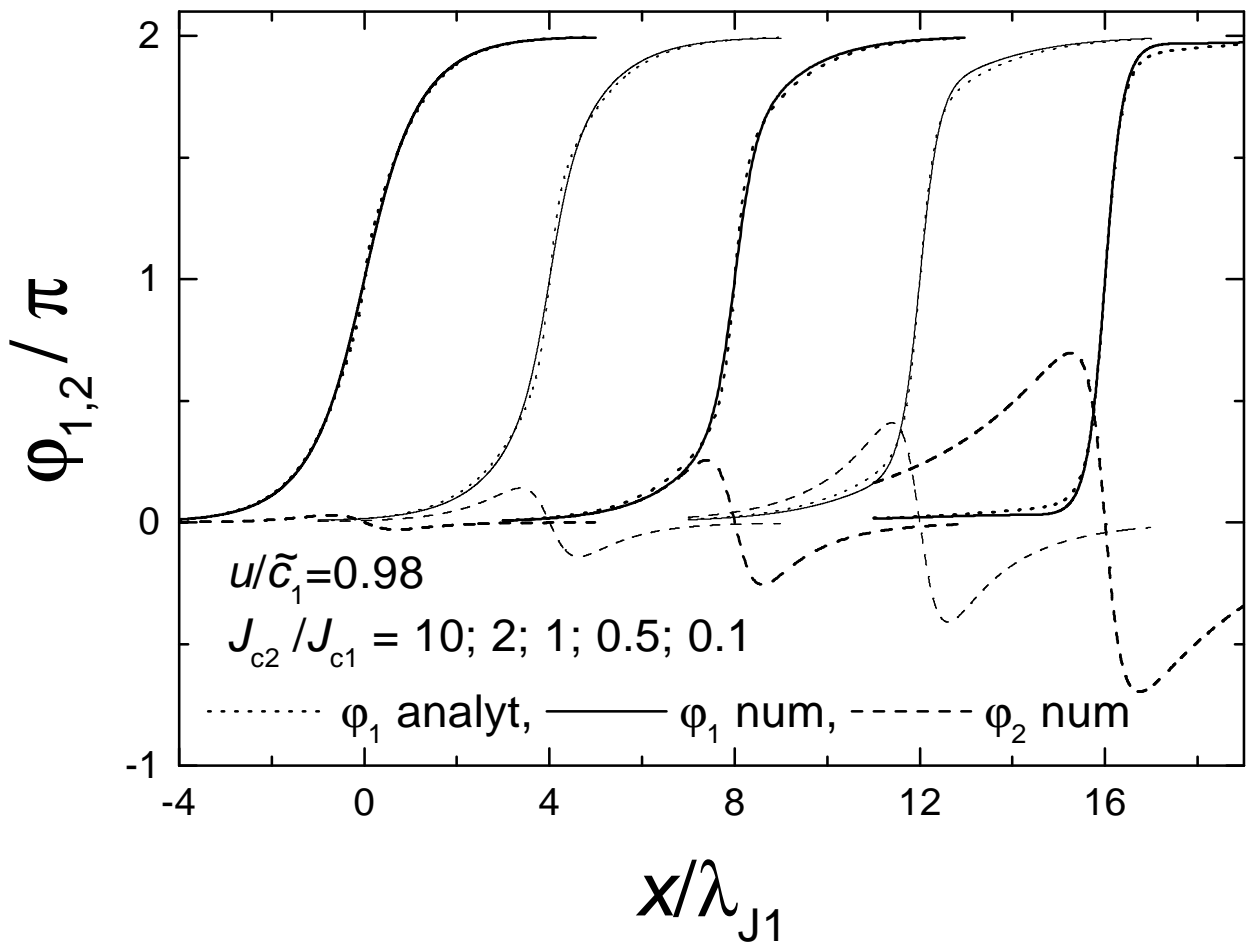
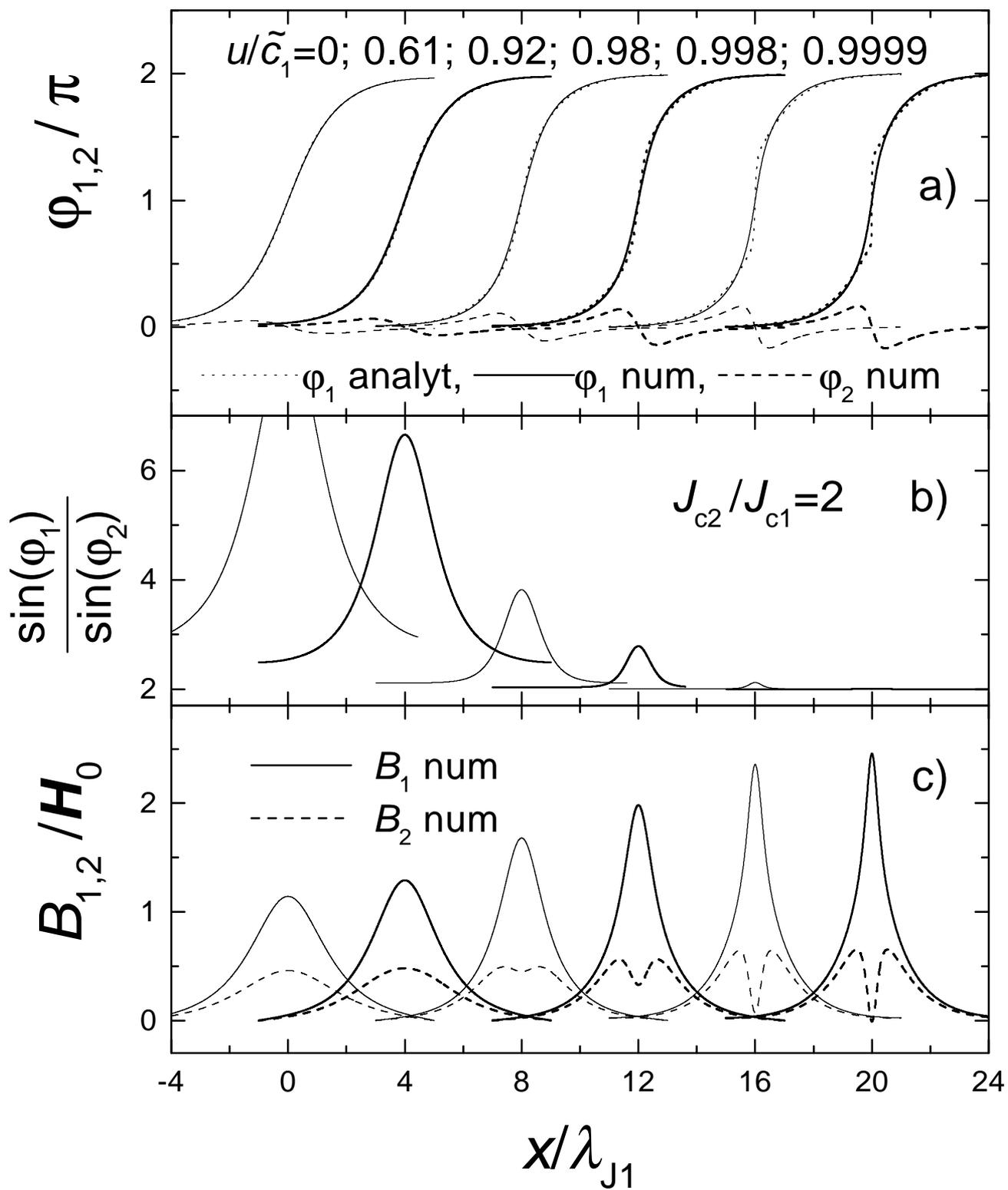
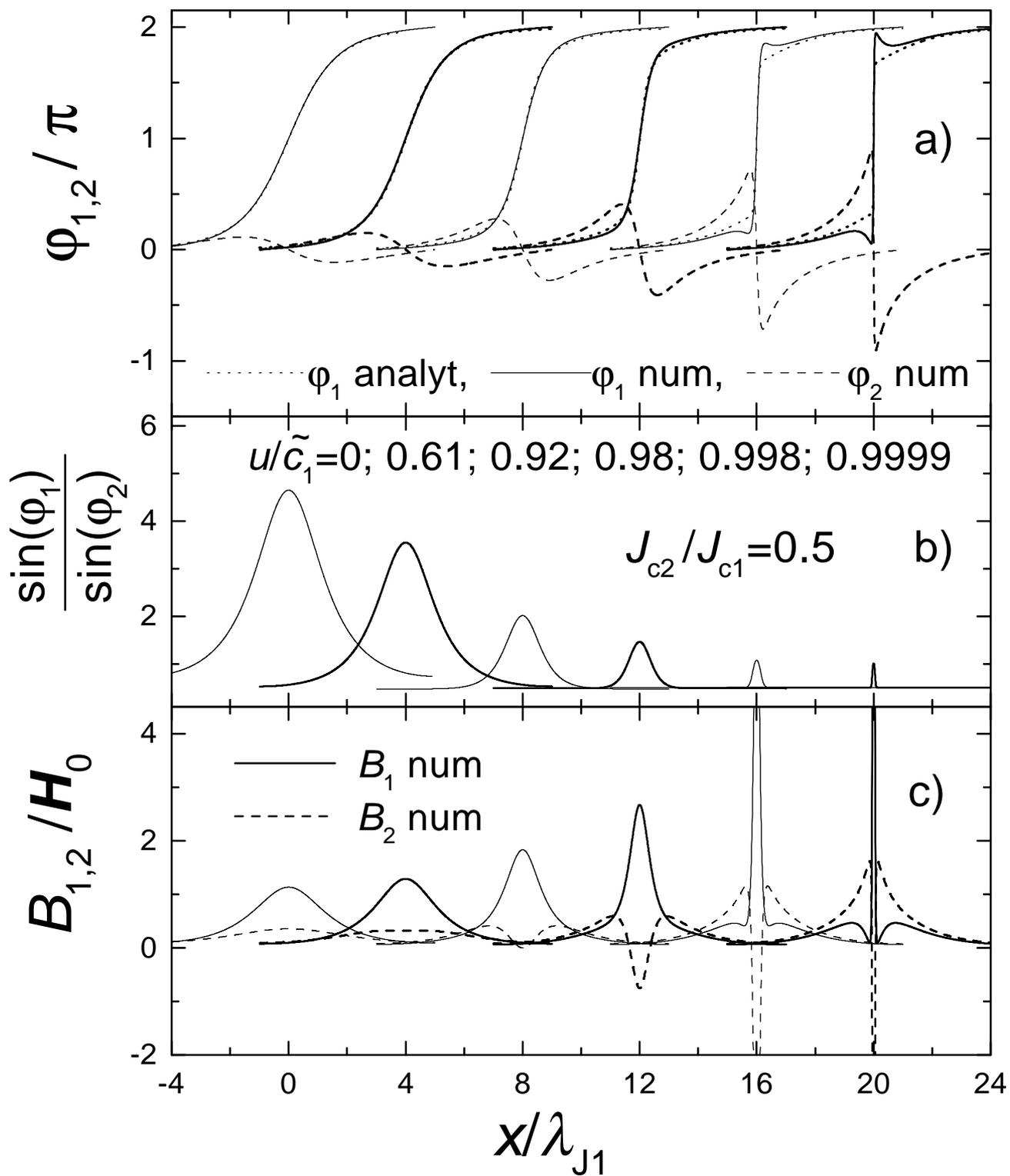
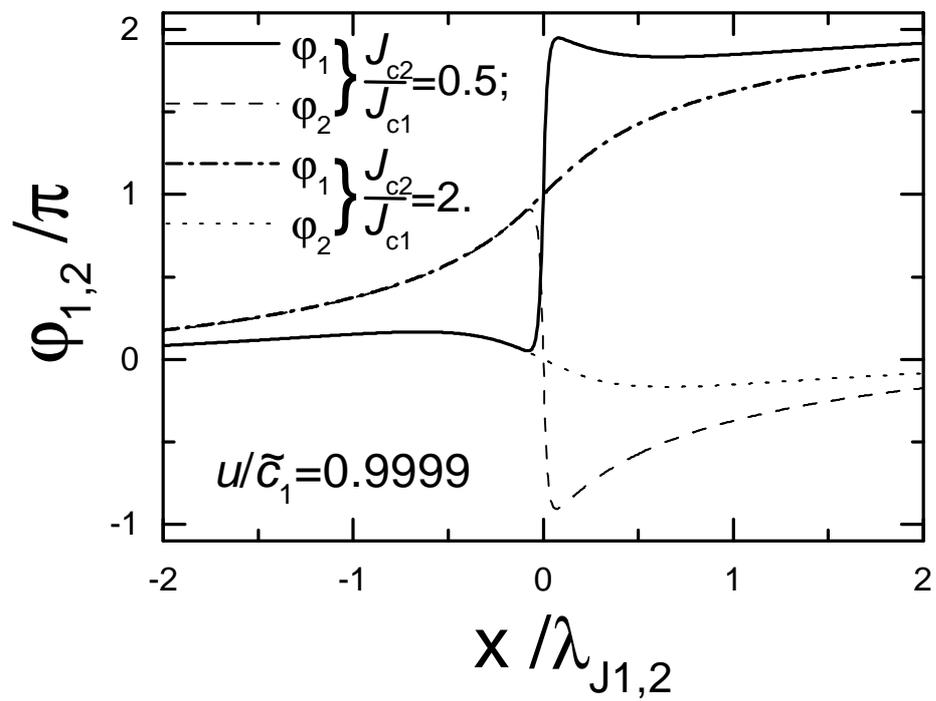
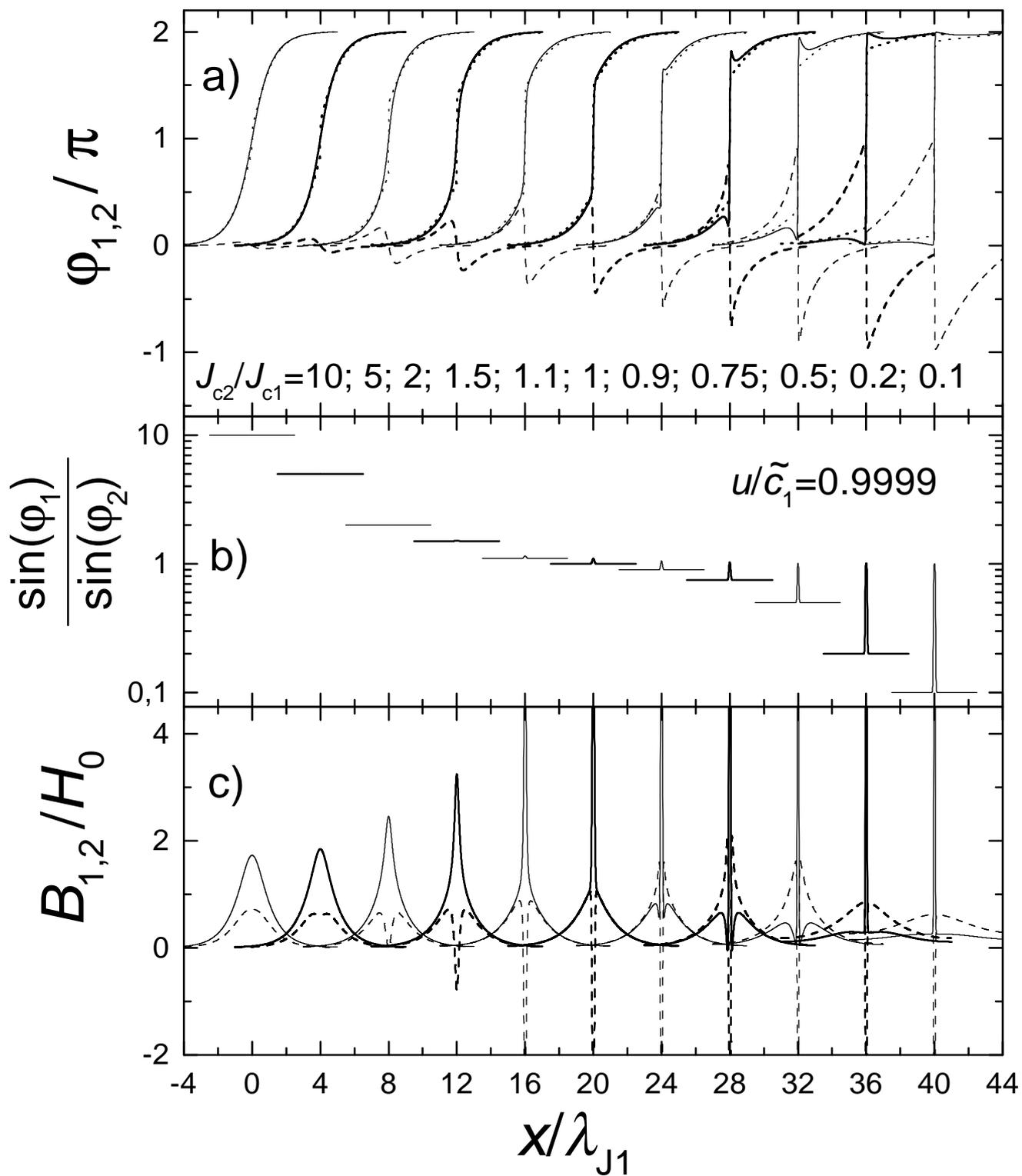
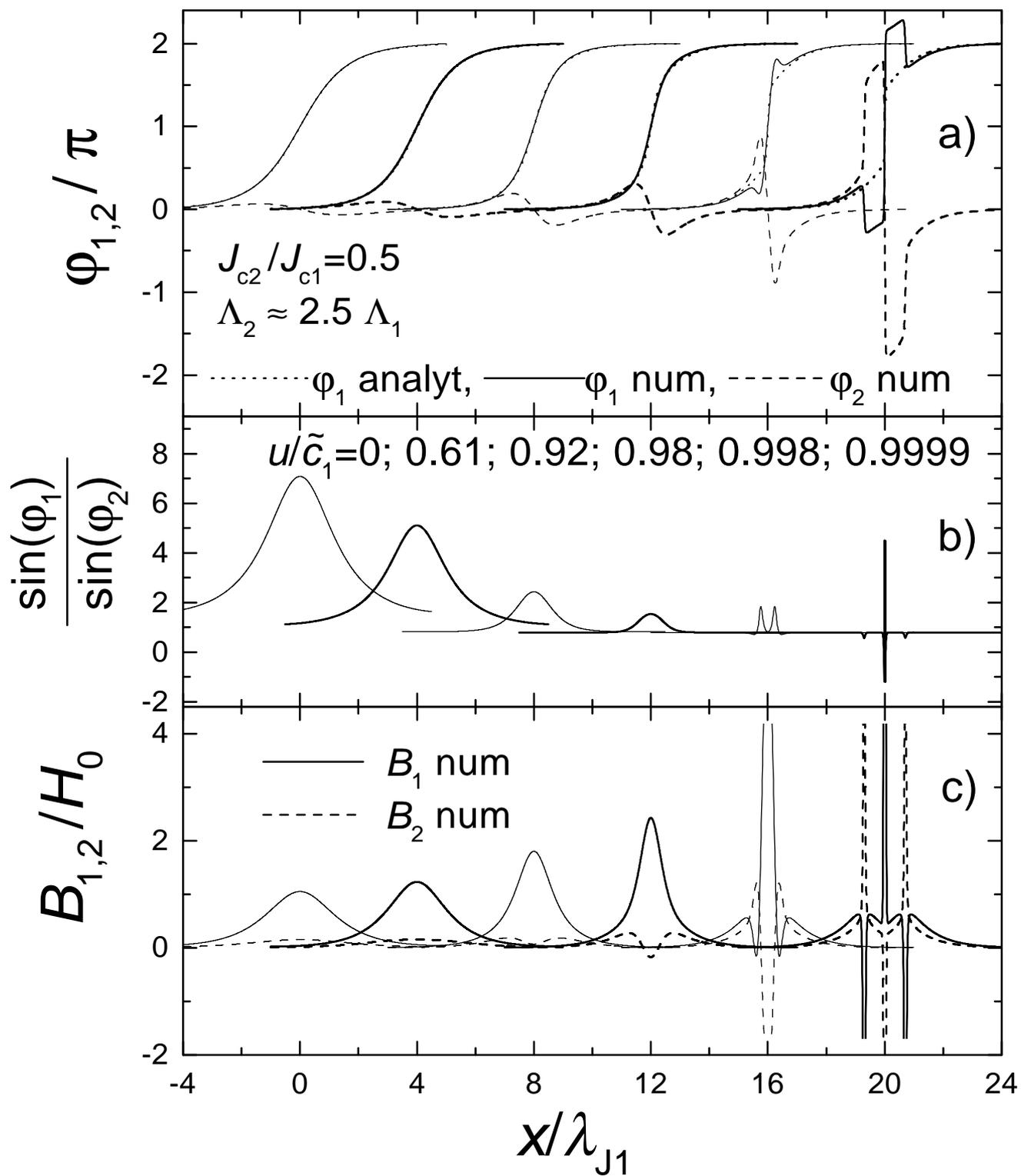
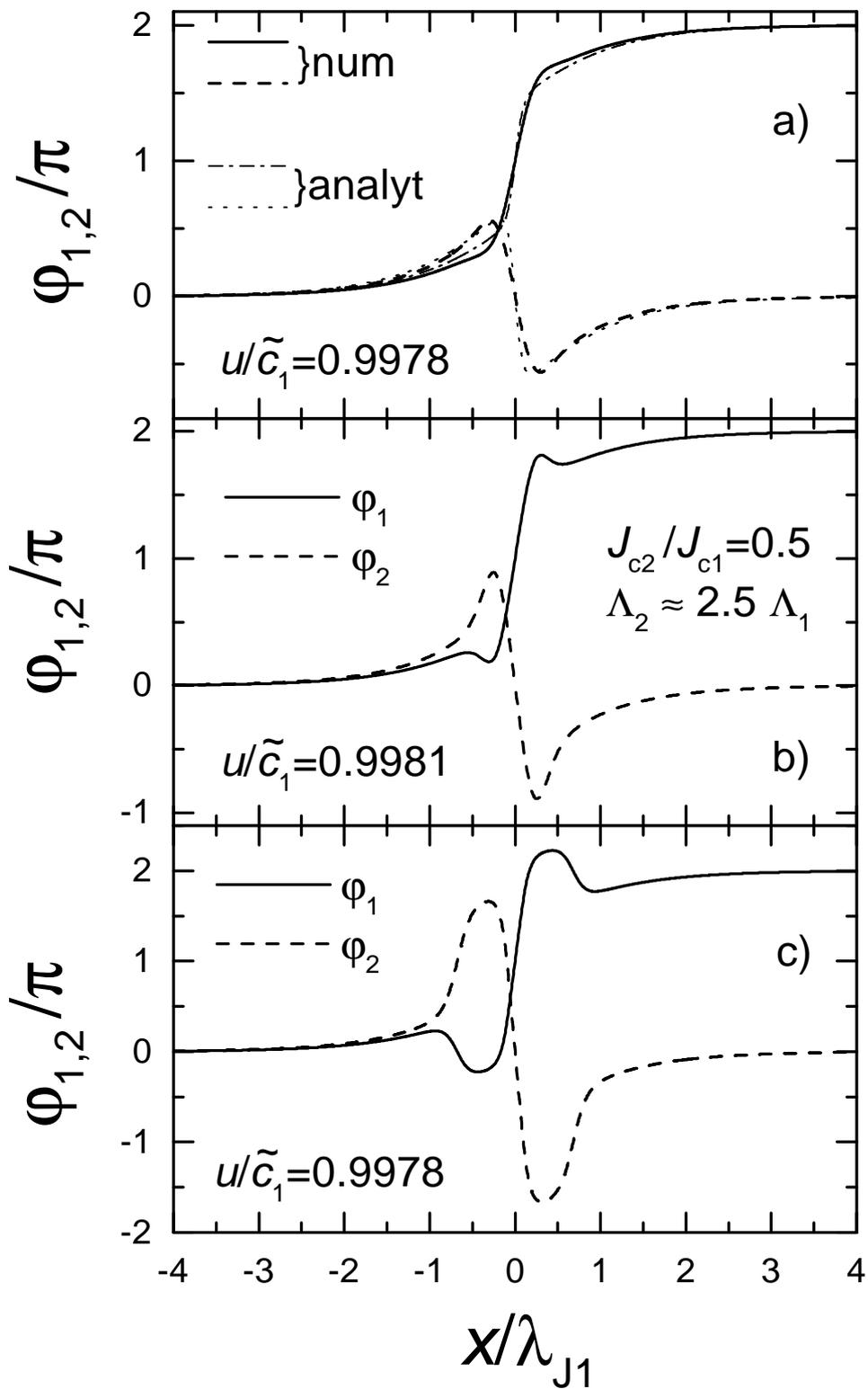
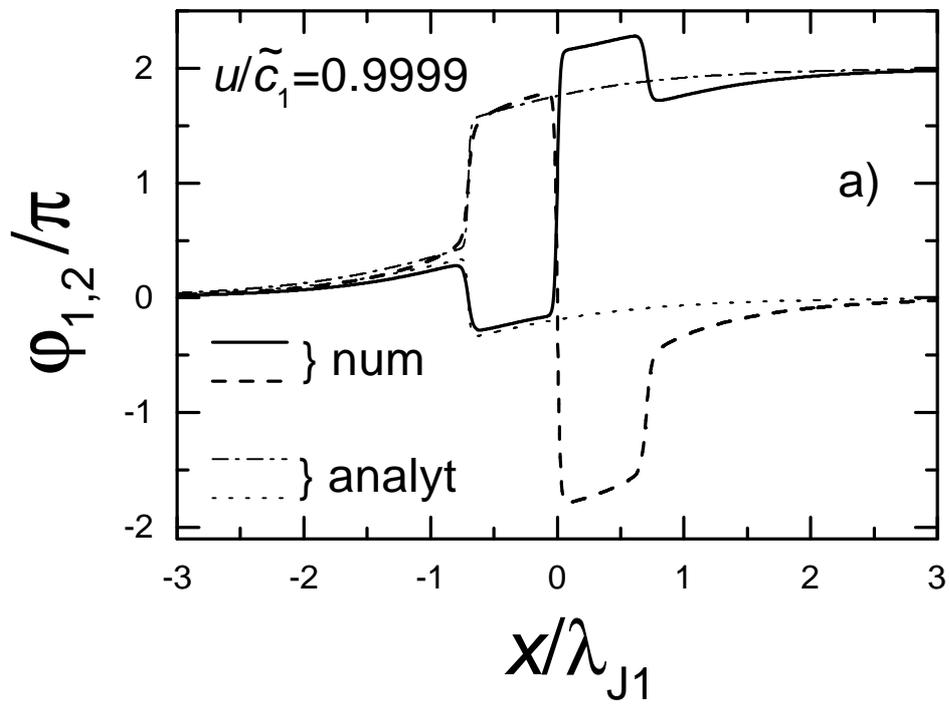
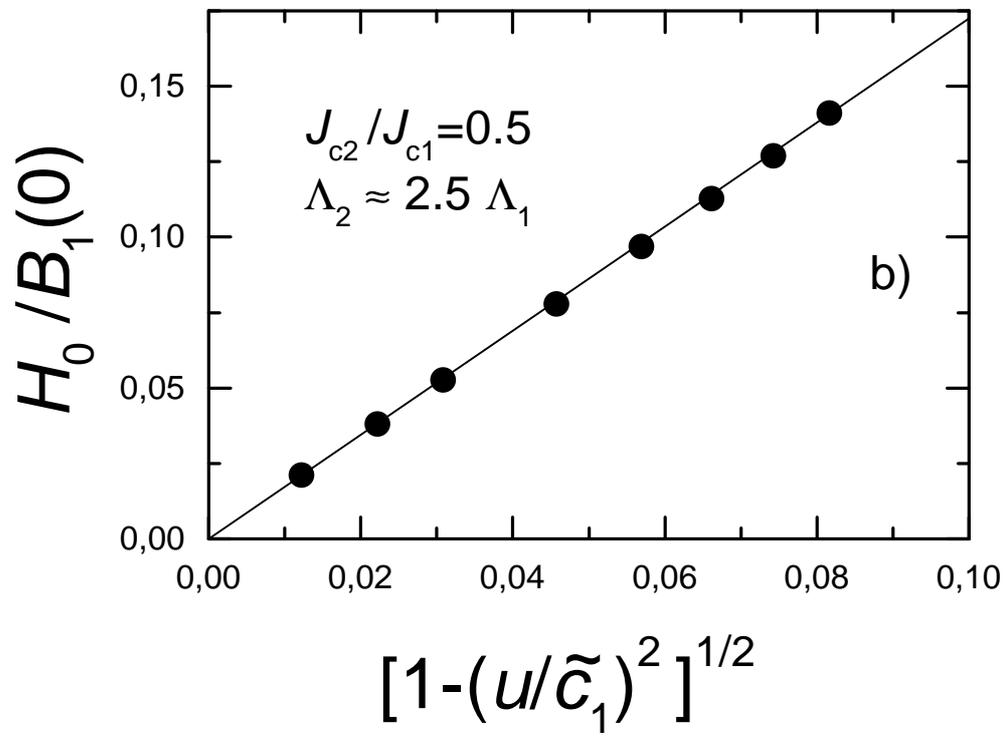
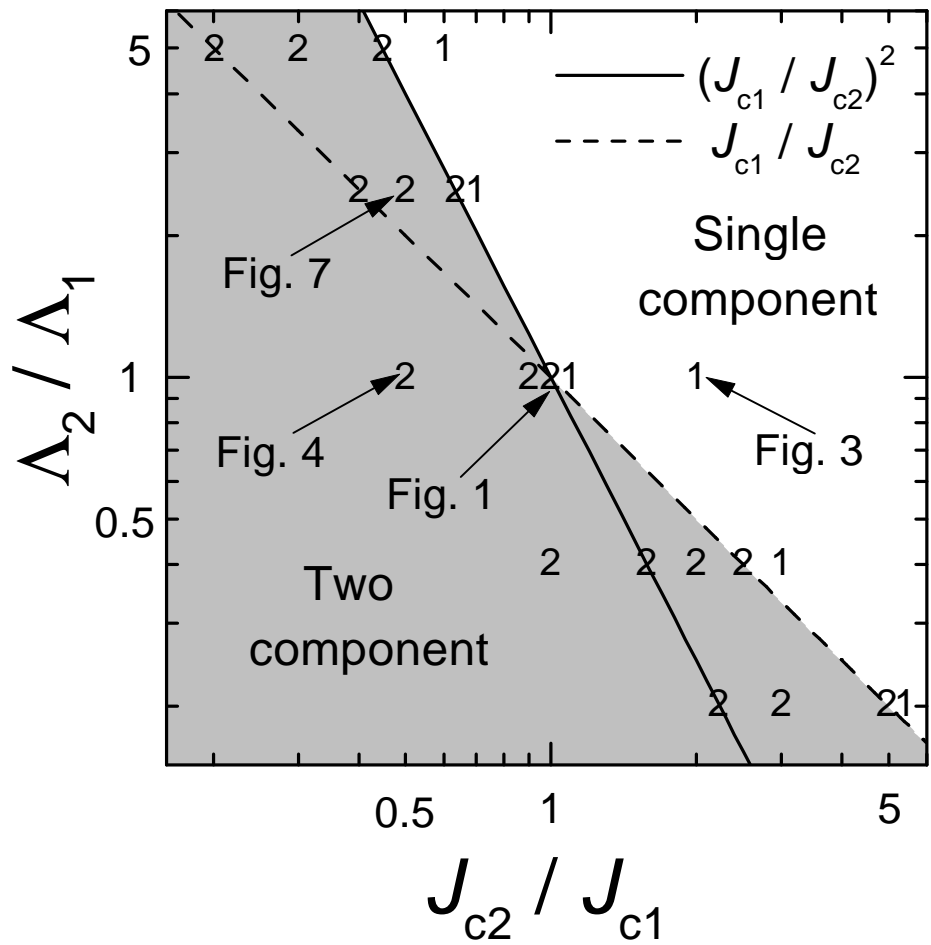
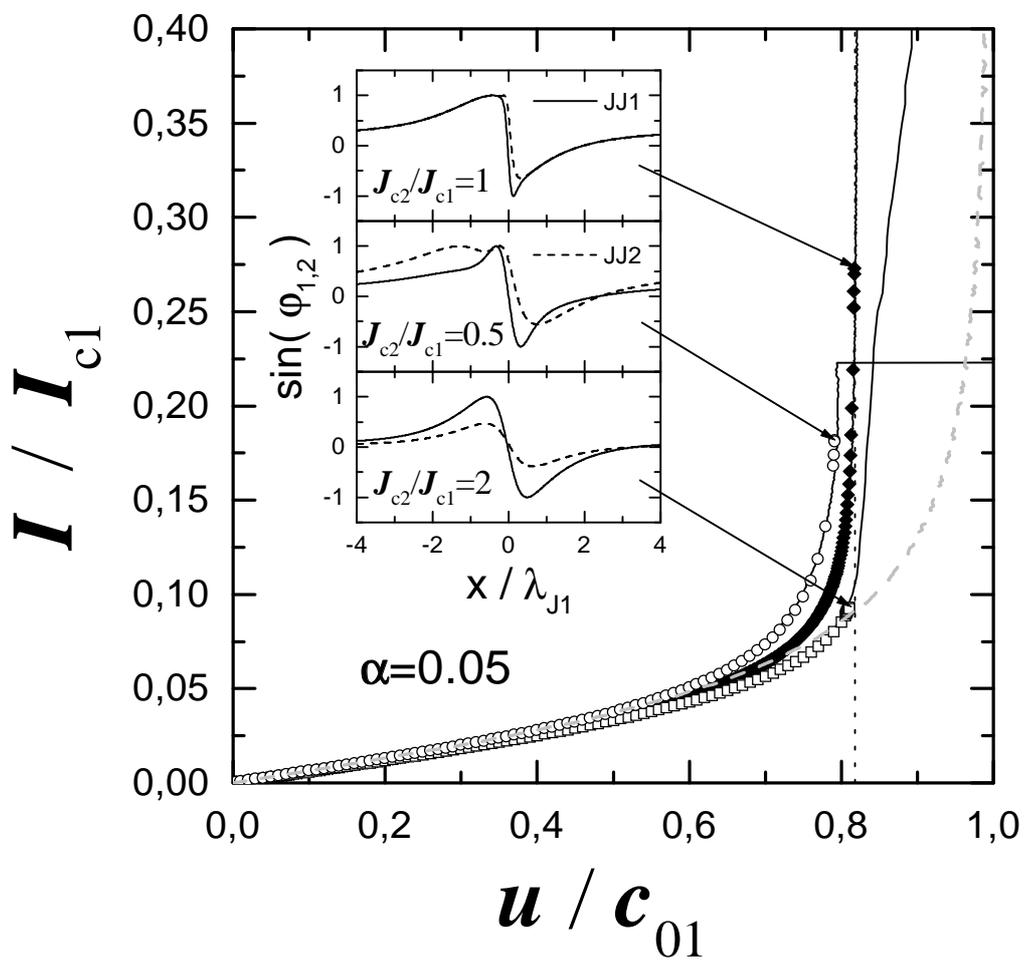
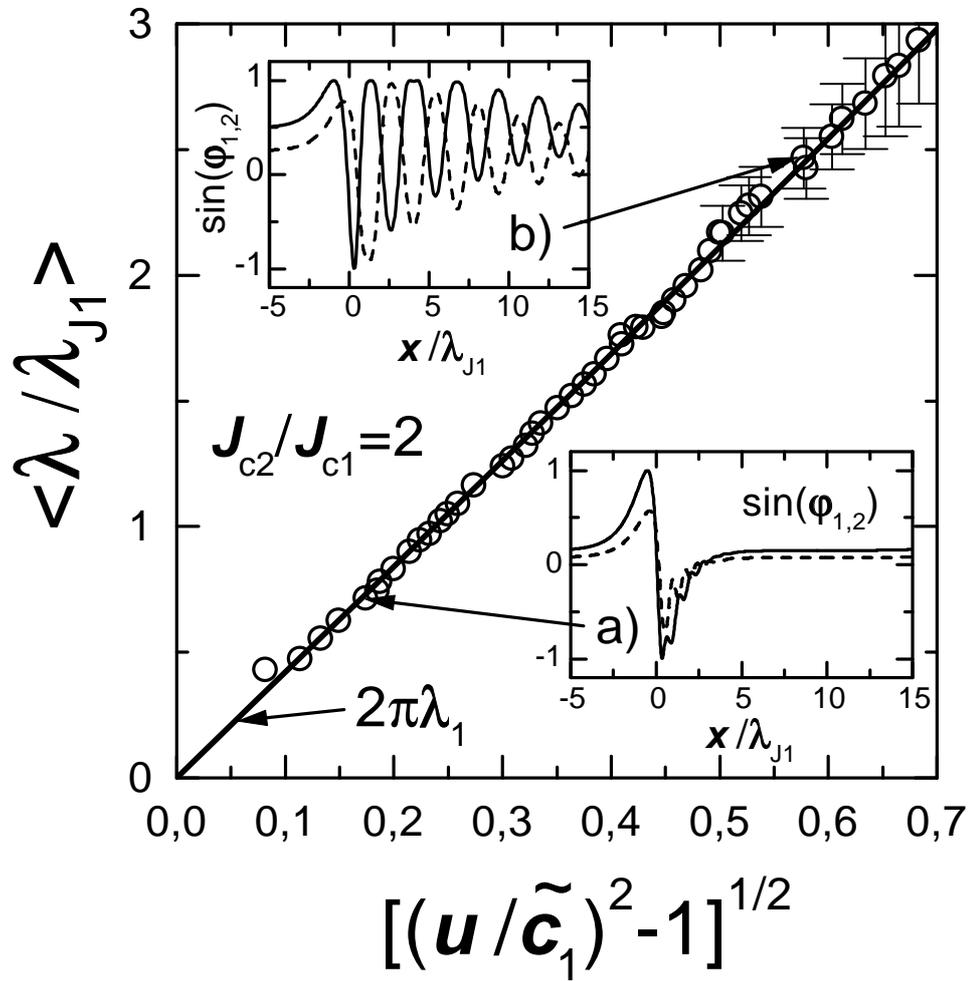
\end{document}